\documentclass[twocolumn,english,prx,superscriptaddress,reprint,showpacs,showkeys,longbibliography]{revtex4-1}
\usepackage[T1]{fontenc}
\usepackage[utf8x]{inputenc}
\usepackage{color}
\usepackage{babel}
\usepackage{units}
\usepackage{amsmath}
\usepackage{amssymb}
\usepackage{graphicx}
\usepackage{esint}
\usepackage[unicode=true,pdfusetitle,
 bookmarks=true,bookmarksnumbered=false,bookmarksopen=false,
 breaklinks=false,pdfborder={0 0 1},backref=false,colorlinks=true]
 {hyperref}
\hypersetup{
 allcolors=blue}

\makeatletter
\usepackage{textgreek}
\usepackage{babel}

\makeatother

\begin{document}
\global\long\def\vect#1{\overrightarrow{\mathbf{#1}}}%
\global\long\def\abs#1{\left|#1\right|}%
\global\long\def\av#1{\left\langle #1\right\rangle }%
\global\long\def\ket#1{\left|#1\right\rangle }%
\global\long\def\bra#1{\left\langle #1\right|}%
\global\long\def\tensorproduct{\otimes}%
\global\long\def\braket#1#2{\left\langle #1\mid#2\right\rangle }%
\global\long\def\omv{\overrightarrow{\Omega}}%
\global\long\def\inf{\infty}%

\title{Breakdown of universality in three-dimensional Dirac semimetals with
random impurities}
\author{J. P. Santos Pires}
\email{up201201453@up.pt}

\address{Centro de Física das Universidades do Minho e Porto, University of
Porto, 4169-007 Porto, Portugal}
\address{Department of Physics, University of Central Florida, Orlando, Florida
32816, USA}
\author{B. Amorim}
\address{Centro de Física das Universidades do Minho e Porto, University of
Minho, 4710-057 Braga, Portugal}
\author{Aires Ferreira}
\email{aires.ferreira@york.ac.uk}

\address{Department of Physics and York Centre for Quantum Technologies, University
of York, York YO10 5DD, United Kingdom}
\author{İnanç Adagideli}
\address{Faculty of Engineering and Natural Sciences, Sabancı University, 34956
Orhanlı-Tuzla, Turkey}
\address{Faculty of Science and Technology and MESA+ Institute for Nanotechnology,
University of Twente, 7500 AE Enschede, Netherlands}
\author{Eduardo R. Mucciolo}
\address{Department of Physics, University of Central Florida, Orlando, FL
32816, USA}
\author{J. M. Viana Parente Lopes}
\email{jlopes@fc.up.pt}

\address{Centro de Física das Universidades do Minho e Porto, University of
Porto, 4169-007 Porto, Portugal}
\begin{abstract}
Dirac-Weyl semimetals are unique three-dimensional (3D) phases of
matter with gapless electrons and novel electrodynamic properties
believed to be robust against weak perturbations. Here, we unveil
the crucial influence of the disorder statistics and impurity diversity
in the stability of incompressible electrons in 3D semimetals. Focusing
on the critical role played by rare impurity configurations, we show
that the abundance of low-energy resonances in the presence of diluted
random potential wells endows rare localized zero-energy modes with
statistical significance, thus lifting the nodal density of states.
The strong nonperturbative effect here reported converts the 3D Dirac-Weyl
semimetal into a compressible metal even at the lowest impurity densities.
Our analytical results are validated by high-resolution real-space
simulations in record-large 3D lattices with up to $536\,000\,000$
orbitals. 
\end{abstract}
\maketitle

\section{Introduction}

The discovery of Dirac and Weyl semimetals (DWSMs) has provided a
rich arena for probing novel gapless phases of matter with unique
transport properties and topological features\,\citep{Armitage2018}.
Several types of gapless systems featuring Dirac or Weyl points in
three-dimensional (3D) momentum space have been realized\,\citep{Borisenko2014,Han-Jin2017,Xu613}.
The simplest DWSMs exhibit twofold or fourfold degenerate linear-band
touching points at the Fermi level with isotropic velocities and a
possible replication into disjoint momentum-space valleys. Their pointlike
Fermi surface is protected against band gap opening due to either
topological constraints---in Weyl systems with broken time-reversal
({\small{}$\mathcal{T}$}) or inversion symmetries ({\small{}$\mathcal{P}$})\,\citep{Armitage2018}
--- or crystal symmetries in {\small{}$\mathcal{TP}$}-symmetric
Dirac systems\,\citep{Young2012,Yang2014}. Thus, any clean DWSM
is an incompressible electron gas with a quadratically vanishing density
of states (DoS). \textcolor{black}{Despite the inefficient charge
screening at the node, this paradigm is believed to survive electron-electron
Coulomb interactions, giving way to a marginal Fermi liquid behavior\,\citep{PhysRevB.90.121107,PhysRevB.92.045104}.
In addition, weakly interacting DWSMs can display strongly renormalized
Fermi velocities, but the nodes' integrity and topology are expected
to remain robust\,\citep{Interact_PhysRevB.94.241102,Interact_PhysRevB.97.161102,Interact_PhysRevB.98.241102,Interact_PhysRevLett.113.136402,Roy16,Roy17,Roy18,maciejko2014}. }

An outstanding question is whether random on-site potentials ubiquitous
in realistic systems (e.g., due to impurities in the crystal lattice)
can give way to a compressible diffusive metallic phase with a finite
nodal DoS\,\citep{Fradkin1986_2,Shindou2009,Goswami2011,Yang2014,Nandkishore2014,Syzranov2015,Sbierski2016,Pixley2016,Pixley2016_2,pixley2016_3,Syzranov2016,Buchhold2018,Buchhold2018_2,Ziegler2018,Goncalves2020,Wilson2020,Kobayashi2020}.
An early result by Fradkin predicted that Dirac nodes are stable in
{\small{}$d\!=\!2\!+\!\epsilon$} dimensions, below some critical
disorder strength\,\citep{Fradkin1986_2}. The robustness of DWSMs
against weak random perturbations is best visualized by considering
a massless particle moving through a short-ranged random potential
of strength{\small{} $W$}. Since, near a node, the de Broglie wavelength,
{\small{}$\lambda\!=\!\hbar v/E$}, largely exceeds the disorder correlation
length, the central limit theorem applies and the fluctuations around
the average potential inside a volume {\small{}$\lambda^{d}$ }must
scale as {\small{}$\delta V\!\!\propto\!W\!\lambda^{\!{\scriptscriptstyle -d/2}}$}.
In{\small{} $d\!=\!3$}, the fluctuations vanish as{\small{} $E^{{\scriptscriptstyle 3/2}}$,}
i.e., faster than the band energy near a node, rendering the semimetal
phase stable.

The early field-theoretical point of view has been recently questioned
by nonperturbative calculations\,\citep{Nandkishore2014, Pixley2016_2},
hinting that 3D gapless phases can become unstable due to the emergence
of\emph{ zero-energy states} bound to statistically rare regions of
the disorder potential landscape. According to this picture, the nodal
DoS remains nonzero for arbitrarily weak disorder without any signature
of singular behavior. Evidence for avoided quantum criticality (AQC)
facilitated by localized nodal eigenstates has been provided by lattice
simulations of a 3D Dirac model with uncorrelated on-site disorder\,\citep{Pixley2016_2}.
Challenging these findings, Buchhold \textit{et al.} noted that rare
events are preceded by scattering resonances which always carry zero
spectral weight at a node. Furthermore, as they are only possible
for fine-tuned (\textit{``magical}'') impurity configurations, this
would imply that the nodal DoS cannot be lifted at variance with the
AQC scenario\,\citep{Buchhold2018,Buchhold2018_2}. Their claim is
backed by the exact solution of a Weyl node with a spherical impurity.\textcolor{blue}{{}
}\textcolor{black}{These paradoxical findings have attracted significant
attention recently\,\citep{Pixley2016_2,roy_universal_2016,Ziegler2018,Wilson2020}
because they question the phase stability of incompressible 3D gapless
phases in realistic conditions. Moreover,}\textit{\textcolor{black}{{}
}}\textcolor{black}{the driving mechanism for semimetal-to-metal transitions
in the phase diagram of dirty 3D DWSMs remains elusive.}

\textcolor{black}{In this paper, we resolve this conundrum by tackling
the spherical impurity problem using two complementary theoretical
approaches. First, within a continuum model, we argue that an unforeseen
non-analytic behavior of scattering phase shifts at the node obstructs
a direct use of Friedel's sum rule (FSR)} \,\textcolor{black}{\citep{Buchhold2018,Buchhold2018_2}.
This difficulty can be overcome by keeping track of the level statistics
in systems of increasingly large volume at fixed impurity concentration.
The puzzling behavior of the phase shifts is explained by the emergence
and sudden disappearance of bound states}\textit{\textcolor{black}{\emph{
upon tuning the impurity potential across a fine-tuned}}}\textit{\textcolor{black}{{}
``magical value}}\textcolor{black}{''. Crucially, our level statistics
analysis reveals that rare bound states are accompanied by a }\textcolor{black}{\emph{continuum
of low-energy resonances}} surrounding the nod\textcolor{black}{e
in realistic material systems with a }\textcolor{black}{\emph{diversity}}\textcolor{black}{{}
of random} short-range impurities.\textcolor{red}{{} }Such near-critical
impurity configurations give effective statistical weight to \textit{\textcolor{black}{magical}}
impurities and ultimately endow the node with a finite average DoS.
Second, we carry out high-precision numerical calculations in a lattice
version of the problem, hosting one or more impurities. Remarkably,
our real-space simulations not only unambiguously demonstrate the
destabilization of a 3D DWSM by a diversity of random near-critical
impurities, but also quantitatively agree with the continuum theory
predictions\textcolor{black}{{} in the dilute impurity regime. These
findings provide the missing link between continuum and lattice approaches
to the DWSM theoretical puzzle and unambiguously pinpoint the newly
unveiled }\textit{\textcolor{black}{statistical significance of near-critical
impurities}}\textcolor{black}{{} as the driving mechanism for AQC. Hence,
dilute impurities can only destabilize a DWSM provided their random
parameters are drawn with a probability density which is nonzero on
(at least) one}\textit{\textcolor{black}{{} magical}}\textcolor{black}{{}
value.} Lastly, we note that subtleties in disordered Dirac systems
have a long history\,\citep{Atkinson2000,Hirschfeld2002, Pepin2001, Adagideli2002, Altland2002}.\,For
instance, in 2D $d$-wave superconductors, there are four low-energy
quasiparticle Dirac valleys, and scalar impurities are pair breaking.
The latter induce resonances that, in the strong scattering limit,
turn into sharp DoS peaks at $E\!\!=\!\!0$ (Majorana zero modes)\,\citep{Balatsky95, Adagideli99}.

\textcolor{black}{\vspace{-0.1cm}
 }

\textcolor{black}{The remainder of the paper is organized as follows:\,In
Sec.\,\ref{sec:Continuum-theory}, we present the theoretical tools
for calculating the DoS correction induced by dilute spherical scalar
impurities hosted within a single-node continuum model of a Dirac
semimetal. We further argue our conclusions to remain valid in 3D
Weyl semimetals. In Sec.\,\ref{sec:Impurity-induced-change-in},
we highlight the main caveats implied by a direct use of FSR and show
how to obtain consistently the thermodynamic limit DoS change due
to a finite (albeit small) concentration of impurities.\,In Sec.\,\ref{sec:Near-critical-impurities-lift}
this theory is used to predict the conditions on which AQC holds in
a Dirac semimetal. Our predictions are validated by high-resolution
real-space calculations in Sec.\,\ref{sec:Lattice-simulations}.
Finally, in Sec.\,\ref{sec:Outlook} we summarize our main findings
and highlight future directions for further study.}

\vspace{-0.3cm}

\section{\label{sec:Continuum-theory}Continuum theory}

\vspace{-0.2cm}

\begin{figure}[t]
\vspace{-0.2cm}

\begin{centering}
\hspace{-0.1cm}\includegraphics[scale=0.165]{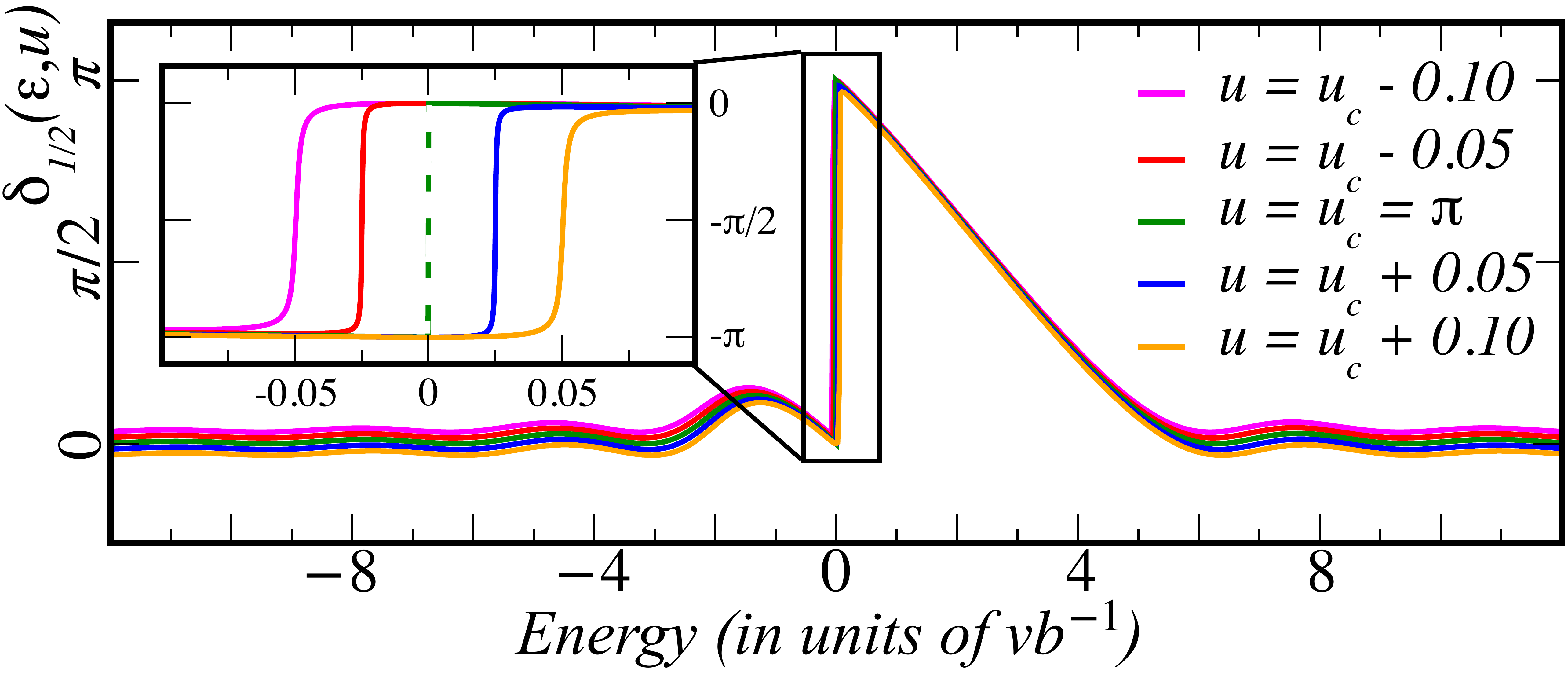} 
\par\end{centering}
\vspace{-0.35cm}

\caption{\label{fig:PhaseShift - -Pi to Pi}Plot of the energy-dependent phase
shift $\delta_{{\scriptscriptstyle 1/2}}\!\left(\varepsilon,u\right)$
in accordance with the prescription $\delta_{j}\!\left(\varepsilon\!\to\!\pm\infty,u\right)\!=\!-u$.
Several values of $u$ are plotted around $u_{c}=\pi$\textcolor{black}{.\,The
inset depicts a close-up of the curves near $\varepsilon\!=\!0$.}\vspace{-0.8cm}
 }
\end{figure}

We start by reviewing the low-energy description o\textcolor{black}{f
a noninteracting single-valley 3D DWSM. The Hamiltonian can be written
as $\mathcal{H}_{0}\!\!=\!v\,\mathbf{\boldsymbol{\alpha}}\!\cdot\mathbf{p}$,
with $\hbar\!\equiv\!1$, $\alpha^{i}\!=\!\sigma^{x}\!\!\otimes\!\!\sigma^{i}$,
$v$ being the Fermi velocity, and $\mathbf{p}\!=\!-\imath\boldsymbol{\nabla}$
being the momentum operator. Here, $\sigma_{i}$ ($i\!\!=\!\!x,\!y,\!z$)
denote Pauli matrices acting on internal spin space. Introducing a
scalar impurity potential in the Hamiltonian breaks translation invariance,
but if $\mathcal{U}\!\left(\mathbf{r}\right)\!\!=\!\mathcal{U}\!\left(\abs{\mathbf{r}}\right)$,
rotational symmetry around the impurity center is preserved. For concreteness,
we consider a spherical well or plateau potential, $\mathcal{U}\!\left(\mathbf{r}\right)\!\!=\!\!\lambda\Theta\left(b\!-\!\left|\mathbf{r}\right|\right)$\,\citep{Nandkishore2014}.
We note that this model is suitable to describe realistic multiple-valley
Dirac or Weyl semimetals insofar as the impurity radius $b$ is much
larger than the lattice spacing (thus effectively suppressing intervalley
scattering). The coupling of distinct Weyl sectors at each valley
is also absent due to the scalar structure of the impurity potential.
The eigenstates of $\mathcal{H}\!=\!\mathcal{H}_{0}\!+\!\mathcal{U}\!\left(\abs{\mathbf{r}}\right)$
can be written as}

\begin{equation}
\Psi_{j,j_{z}}^{\kappa}\!\left(\mathbf{r}\right)\!=\!\left[\frac{f_{j}^{\kappa}\!\left(r\right)}{r}\Theta_{j,j_{z}}^{-\kappa}\!\left(\hat{\mathbf{r}}\right),\frac{\iota g_{j}^{\kappa}\!\left(r\right)}{r}\Theta_{j,j_{z}}^{\kappa}\!\!\left(\hat{\mathbf{r}}\right)\right]^{\text{T}},\label{eq:SpinorAngularStructure-1}
\end{equation}
where $j\in\{\nicefrac{1}{2},\nicefrac{3}{2},\cdots\}$ and $j_{z}\in\left\{ -j,-j+1,\cdots,j\right\} $
are the total angular momentum quantum numbers, while $\kappa=\pm1$
labels the eigenvalues of $\mathcal{K}\!=\!\gamma^{0}\!\cdot\!(2\mathbf{L}\cdot\mathbf{S}\!-\!1)$,
i.e., $\kappa(j\!+\!\nicefrac{1}{2})$. Furthermore, $\Theta_{j,j_{z}}^{\pm}\!\!\left(\hat{\mathbf{r}}\right)$
are orthonormal spin-$\nicefrac{1}{2}$ spherical harmonics and $f_{j}^{\kappa}\left(r\right)$/$g_{j}^{\kappa}\left(r\right)$
are radial functions. For nonzero energies, the latter are radial
spherical waves with phase shifts introduced by the central potential
(see Appendix\,\eqref{sec:AppendixA} for additional details). In
each $j$ sector, the scattering phase shifts $\delta_{j}$ induced
by a spherical well or plateau are obtained by constraining the spinor
to be continuous at $r\!=\!b$. One obtains, after a somewhat lengthy
calculation, 
\begin{widetext}
\vspace{-0.4cm}

\begin{equation}
\tan\delta_{j}\left(\varepsilon,u\right)=\frac{\text{sgn}\left(\varepsilon\!-\!u\right)J_{j+1}\left(\abs{\varepsilon}\right)J_{j}\left(\abs{\varepsilon-u}\right)-\text{sgn}\left(\varepsilon\right)J_{j}\left(\abs{\varepsilon}\right)J_{j+1}\left(\abs{\varepsilon-u}\right)}{\text{sgn}\left(\varepsilon\right)\text{sgn}\left(\varepsilon-u\right)Y_{j+1}\left(\abs{\varepsilon}\right)J_{j}\left(\abs{\varepsilon-u}\right)-Y_{j}\left(\abs{\varepsilon}\right)J_{j+1}\left(\abs{\varepsilon\!-\!u}\right)},\label{eq:B/A-1-1-1}
\end{equation}

\vspace{-0.35cm}
 
\end{widetext}

\noindent where $u\!\equiv\!\lambda b/v_{\text{F}}$, $\varepsilon\!\equiv\!Eb/v_{\text{F}}\!\neq\!(0,u)$,
and $J_{n}(x)\,[Y_{n}(x)]$ are Bessel functions of the first (second)
kind. We underline that Eq.\,(\ref{eq:B/A-1-1-1}) is equivalent
to that obtained in Refs.\,\citep{Nandkishore2014,Buchhold2018_2}
for the Weyl equation. This is unsurprising because our $4\times4$
Dirac model is gapless and the impurity potential has a scalar structure
(i.e. distinct Weyl sectors at each valley remain decoupled). However,
Eq.\,\eqref{eq:B/A-1-1-1} only defines $\delta_{j}\!\left(\varepsilon,u\right)$\,mod$\pi$.
The ambiguity corresponds, at most, to a global change in the sign
of the wavefunction. In order to obtain a unique definition of $\delta_{j}\left(\varepsilon,u\right)$,
one needs to choose a reference point, i.e., as the potential is switched
off ($u\!\to\!0$), the require that the phase shifts vanish across
the entire spectrum\textcolor{black}{. A way to guarantee this is
to enforce that $\delta_{j}\left(\varepsilon\!\to\!\pm\infty,u\right)\!=\!-u$\,\citep{Ma85,Ma_2006}\,}\footnote{\textcolor{black}{As $\varepsilon\!\rightarrow\!\pm\infty$, one has
$\tan\delta_{j}\left(\varepsilon,u\right)\!\rightarrow\!-\tan u$.
This differs from the limiting behavior introduced in Ref.~\citep{Ma85},
$\lim\!{}_{\varepsilon\to\pm\infty}\!\delta_{j}\left(\varepsilon,u\right)\!\!=\!\!\mp u$.
The difference comes from the $\text{sgn}(\varepsilon)$ factors introduced
when defining the radial scattering wavefunctions {[}see Eqs.\,\eqref{eq:g+FreeSolution-3-2}---\eqref{eq:g-FreeSolutionOutside}
of Appendix\,\ref{sec:AppendixA}{]}}}\textcolor{black}{, which is achieved by a trick explained in Appendix\,\ref{sec:AppendixC}.
Since the analysis is qualitatively similar in all $j$ sectors, in
what follows we focus on the $\delta_{1/2}\!\left(\varepsilon\right)$
phase shift (see Fig.\,\ref{fig:PhaseShift - -Pi to Pi}) using the
previous convention (for completeness, plots for other $j$'s and
around different magical $u$'s are provided in Appendix\,\ref{sec:AppendixC}).
For $u\!\!=\!\!u_{c}^{j}$, the phase shift is shown to have a physical
$\pi$ discontinuity at $\varepsilon\!\!=\!\!0$, which marks the
occurrence of zero-energy bound states}\,\citep{Ma_2006}\textcolor{black}{.
For} the 3D massless Dirac equation, bound states at $\varepsilon\!=\!0$
can appear, for particular wells or plateaus, whenever a decoupling
of the radial equations for $f_{j}^{\pm}\left(r\!>\!b\right)/g^{\pm}\left(r\!>\!b\right)$
occurs\,\citep{Nandkishore2014}. In this case, the admissible (asymptotically
decreasing) solutions are simple power laws, $g_{j,k}^{+}\left(r\!>\!b\right)/f_{j,k}^{-}\left(r\!>\!b\right)\!=\!\mathcal{B}^{\pm}r^{-j-1/2}$
and $f_{j,k}^{+}\left(r\!>\!b\right)/g_{j,k}^{-}\left(r\!>\!b\right)\!\!=\!\!0$.
As shown in Appendix\,\eqref{sec:AppendixA}, such spinors are only
continuous at $r\!=\!b$, if the potential satisfies $J_{j}\left(\abs u\right)\!=\!0$.
Hence zero-energy states are allowed in a single-impurity Dirac problem
provided the parameters are fine-tuned, i.e., $\abs{\lambda b/v}\!=\!u_{c}^{j}$
is a root of $J_{j}\left(x\right)$. The critical parameters $\{u_{c}^{j}\}$
are dubbed \emph{magical values}, as they would correspond to rare
regions in a disordered landscape where nonperturbative zero-energy
modes are possible\,\citep{Nandkishore2014,Pixley2016_2}. Note that
these are true (squared-normalizable) impurity bound states within
the Dirac continuum and they have a degeneracy of $2\!\left(\!2j\!+\!1\!\right)$.
These bound-states manifest themselves as a $\pi$ discontinuity at
$\varepsilon\!=\!0$ in the phase shifts when the parameter $u$ crosses
a critical value of that angular momentum channel (inset of Fig.\,\ref{fig:PhaseShift - -Pi to Pi}).
This is in accord with Levinson's theorem for Dirac particles\,\citep{Ma85,Ma_2006}
which states that given an appropriate convention, the number of bound
states is encapsulated in discontinuous $\pi$ jumps of the phase
shifts at zero momentum.


\vspace{-0.7cm}

\section{\label{sec:Impurity-induced-change-in}Impurity-induced change in
the DoS\textup{ }}

\vspace{-0.2cm}

\noindent The change in the DoS induced by an isolated impurity is
conventionally calculated using FSR,

\vspace{-0.55cm}

\begin{equation}
\Delta\nu(\varepsilon,u)=\frac{2}{\pi}\sum_{j=1/2}^{\infty}\left(2j+1\right)\frac{\partial\delta_{j}(\varepsilon,u)}{\partial\varepsilon},\label{eq:FSR}
\end{equation}

\textcolor{black}{\vspace{-0.2cm}
 }

\begin{figure}[t]
\vspace{-0.3cm}

\begin{centering}
\hspace{-0.25cm}\includegraphics[scale=0.235]{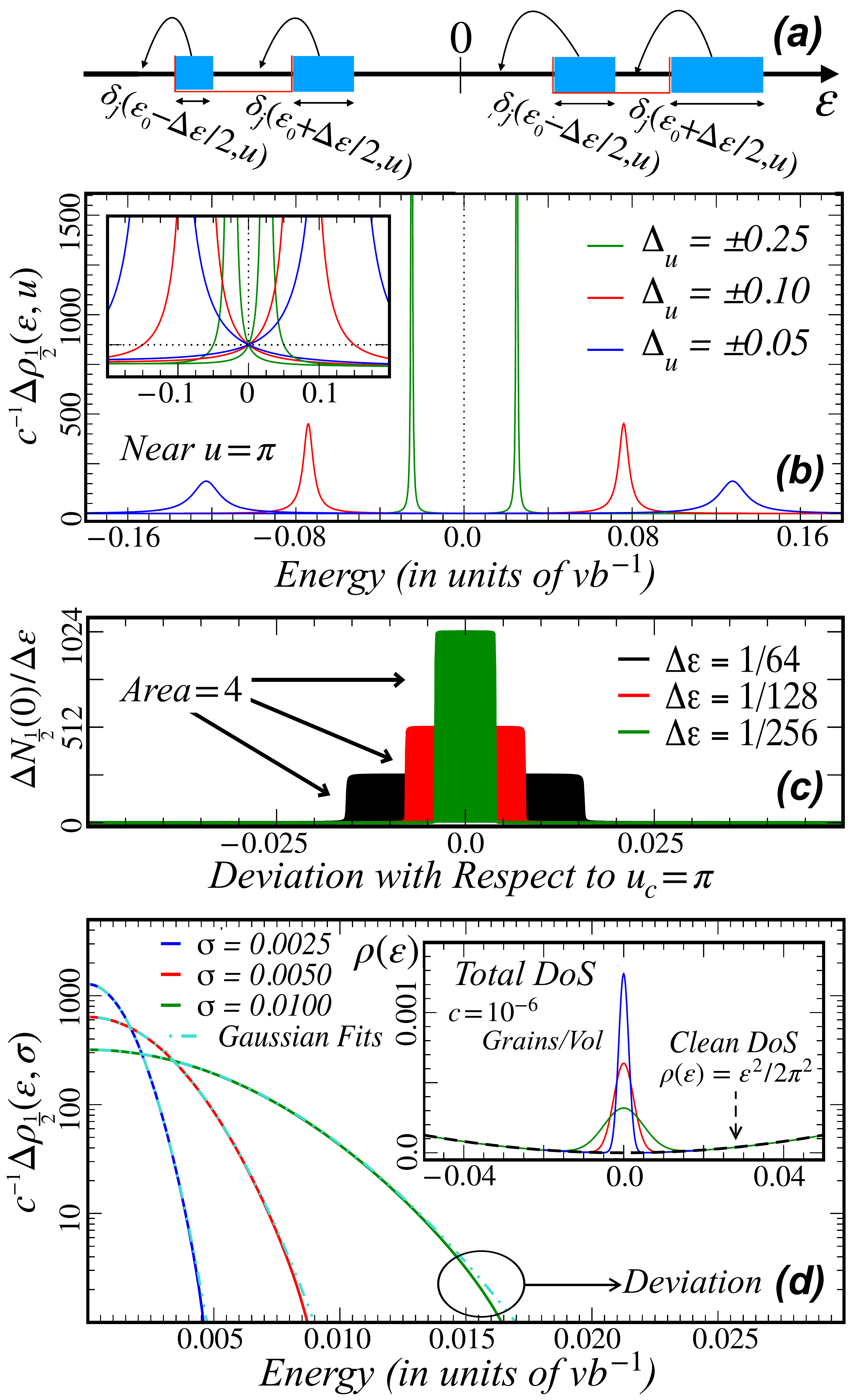}
\par\end{centering}
\vspace{-0.35cm}

\caption{\label{fig:LowEnergyDoS_Altland}(a) Motion of energy levels triggered
by the central impurity. (b) Plots of $\Delta\rho_{{\scriptscriptstyle 1/2}}\!\left(\varepsilon,u\right)$
for selected values of $u$ around $u_{\text{c}}\!=\!\pi$. The curves
were obtained using Eq.(\ref{eq:FSR}) with the phase shifts represented
in Fig.\ref{fig:PhaseShift - -Pi to Pi}. As one approaches $u_{\text{c}}$,
$\Delta\rho_{1/2}\!\left(\varepsilon\right)$ takes the form of a
low-lying peak, which gets narrower and closer to $\varepsilon\!=\!0$
(roughly conserving the area between zeros). The inset shows that
$\Delta\rho_{1/2}\!\left(\varepsilon,u\right)$ is always equal to
zero at $\varepsilon\!=\!0$ for any non-critical $u$. (c) Depiction
of $\Delta N_{1/2}\!\left(\varepsilon\!=\!0,\pi\right)/\Delta\varepsilon$
converging towards a distribution $4\delta\!\left(u\!-\!\pi\right)$
as $\Delta\varepsilon\!\to\!0^{+}$. (d) Theoretical prediction for
$\Delta\rho_{1/2}\!\left(\varepsilon\right)$ due to dilute spherical
impurities with a Gaussian diversity of width $\sigma$ around $u_{c}^{{\scriptscriptstyle \!1\!/\!2}}\!\!=\!\pi$.
The inset depicts the total DoS for $10^{-6}b^{-1}$ impurities per
volume. \vspace{-0.8cm}
}
\end{figure}

\noindent \textcolor{black}{which measures the variation in the number
of states per unit energy (an extensive quantity, not to be confused
with $\Delta\nu$ per unit volume hereafter denoted by $\Delta\rho$).
Strik}ingly, the phase-shift discontinuity caused by the impurity
bound state in the 3D DWSM problem precludes the direct use of FSR,
a peculiar effect that has gone unnoticed in earlier studies. Therefore,
to determine $\Delta\nu$, we adopt a strategy based on counting states
within a finite energy window adapted from Friedel's original reasoning\,\citep{Friedel1952}.
First, we restrict the Dirac fermions to lie inside a finite sphere
of radius $R$, $S_{R}$. The Hermiticity of $\mathcal{H}$ is guaranteed
if the Hilbert space is restricted to states with vanishing current
across the spherical surface, $\partial S_{R}$, that is, to a subspace
where any two spinors $\Psi$ and $\text{\ensuremath{\Phi}}$, satisfy
$\varoiint_{\partial S_{R}}\!d\mathbf{S}\!\cdot\!\left[\!\Psi_{\mu}^{\dagger}\left(\mathbf{r}\right)\!\boldsymbol{\alpha}_{\mu\nu}\Phi_{\nu}\left(\mathbf{r}\right)\!\right]\!=\!0.$
In Eq.\,\eqref{eq:SpinorAngularStructure-1}, this is true if $\cos\!\delta_{j}\left(\varepsilon,u\right)J_{j}\left(\abs{\varepsilon}\!R\right)\!-\!\sin\!\delta_{j}\left(\varepsilon,u\right)Y_{j}\left(\abs{\varepsilon}\!R\right)\!=\!0.$
For each angular momentum sector, this condition quantizes the allowed
energy levels, which we denote by $\varepsilon_{n}^{j}$. The number
of ($j$-sector) levels inside the energy window $[\varepsilon_{0}-\Delta\varepsilon/2,\varepsilon_{0}+\Delta\varepsilon/2]$
is changed by the impurity due to an inwards or outwards migration
of levels from regions of width $\simeq\delta_{j}\left(\varepsilon\pm\nicefrac{\Delta\varepsilon}{2},u\right)/R$
{[}up to $\mathcal{O}\left(R^{-2}\right)${]} near the respective
boundaries. This mechanism is illustrated in Fig. \ref{fig:LowEnergyDoS_Altland}(a).
The variation in the number of $j$ states inside the probing window
is

\vspace{-0.45cm}

\begin{align}
\Delta N_{j}\!\left({\scriptstyle \varepsilon_{0},\Delta\varepsilon,u}\!\right) & \!=\!\frac{2\left(2j\!+\!1\right)}{\pi}\!\left[\delta_{j}\!\left({\scriptstyle \varepsilon_{0}+\frac{\Delta\varepsilon}{2},u}\right)\!-\!\delta_{j}\left({\scriptstyle \varepsilon_{0}-\frac{\Delta\varepsilon}{2},u}\right)\right]\!.\label{eq:Number of States Variation}
\end{align}

\vspace{-0.2cm}

\noindent For a finite $\Delta\varepsilon$, Eq.\,\eqref{eq:Number of States Variation}
is accurate in the asymptotic limit $R\gg1$ and for $\varepsilon_{0}\pm\nicefrac{\Delta\varepsilon}{2}\neq0$
as explained and illustrated in Appendix\,\eqref{sec:AppendixB}.

Next, we consider the intensi\textcolor{black}{ve DoS ($\Delta\rho$)
indu}ced by a finite\textcolor{black}{{} density} ($c$) of impurities
in a volume $V$ focusing on \textcolor{black}{single scattering events.
The neglect of quantum-coherent scattering by multiple impurities
is justified in the dilute regime, where quantum interference corrections
are suppressed by a factor of $1/(k_{F}l)$\,\citep{altland_condensed_2010},
where $l\!\propto\!c^{-1}$ is the mean free path and $k_{F}=E/v\hbar$
is the Fermi wavevector. This is an important assumption confirmed
precisely by our numerical simulations below.}\textcolor{red}{{} }Formally,
the DoS is obtained by the limiting procedure, $\Delta\rho\left(\varepsilon_{0}\right)\!=\!\lim_{\Delta\varepsilon\to0^{+}}\lim_{V\rightarrow\infty}\left[\Delta N\left(\varepsilon_{0},\Delta\varepsilon,\left\{ u_{i}\right\} ,V\right)/V\Delta\varepsilon\right],$
where $i$ indexes the impurity and $\Delta N\left(\varepsilon_{0},\Delta\varepsilon,\left\{ u_{i}\right\} ,V\right)$
is the variation in the total number of states. Assuming that $\left\{ u_{i}\right\} $
are drawn from a probability density function $p\left(u\right)$,
the thermodynamic limit then reads

\vspace{-0.45cm}

\begin{align}
\sum_{i}\frac{\Delta N_{j}(\varepsilon_{0},\Delta\varepsilon,u_{i})}{V}\!\!\underset{{\scriptscriptstyle V\to\infty}}{\longrightarrow}\!c\!\int\!du\,p(u)\Delta N_{j}(\varepsilon_{0},\Delta\varepsilon,u),\label{eq:ThermodynamicLimitDiversity}
\end{align}
\vspace{-0.45cm}

\noindent where $\Delta N_{j}\!\left(\varepsilon_{0},\Delta\varepsilon,u\right)$
is given by Eq.~(\ref{eq:Number of States Variation}). The final
expression for the DoS variation due to a dilute set of random impurities,
$\Delta\rho\left(\varepsilon_{0}\right)\!=\!\!\sum_{j}\Delta\rho_{j}\left(\varepsilon_{0}\right)$,
is obtained from

\vspace{-0.45cm}

\begin{multline}
\Delta\rho_{j}(\varepsilon_{0})\!=\!c\!\lim_{{\scriptscriptstyle \Delta\varepsilon\to0^{+}}}\!\!\int du\,p\left(u\right)\frac{\Delta N_{j}\left(\varepsilon_{0},\Delta\varepsilon,u\right)}{\Delta\varepsilon}.\label{eq:Number of States Variation-2}
\end{multline}

\vspace{-0.3cm}

\noindent 

\noindent 

\noindent The order of limits in Eqs. (\ref{eq:ThermodynamicLimitDiversity})
and (\ref{eq:Number of States Variation-2}) is essential. The integration
over $u$ must be done prior to taking the $\Delta\varepsilon\!\to\!0^{{\scriptscriptstyle +}}$
limit\textcolor{black}{. This is reminiscent of lattice simulations,
where the resolution parameter must be sent to zero only after the
thermodynamic limit has been taken (see below). }If $\delta_{j}\!\left(\varepsilon,u\right)$
is differentiable at $\varepsilon\!=\!\varepsilon_{0}$, the $\Delta\varepsilon\!\to\!0^{{\scriptscriptstyle +}}$
limit can be safely brought inside the integral, and one obtains $\Delta\rho_{j}\left(\varepsilon_{0}\right)\!=\!c\left(4j\!+\!2\right)\!\av{\partial\delta_{j}\left(\varepsilon,u\right)\!/\!\partial\varepsilon|_{\varepsilon=\varepsilon_{0}}\!}_{\!u}\!/\pi,$
where $\langle f\rangle_{u}\!=\!\int\!du\,p\left(u\right)\!f(u)$,
i.e., the familiar FSR. A direct application of FSR\textcolor{black}{\,{[}Eq.\,\eqref{eq:FSR}{]}}
was employed in Refs.\,\citep{Buchhold2018,Buchhold2018_2} to determine
the DoS in Weyl systems with statistical fluctuations of $u$ around
a critical value $u_{c}$, leading to the conclusion that $\Delta\rho_{j}\!\!\left(0\right)\!=\!0$.
This was inferred from the fact that $\partial\delta_{j}\left(\varepsilon,u\right)/\partial\varepsilon|_{\varepsilon=0}\!=\!0$
for any $u\!\neq\!u_{c}^{j}$; see Fig.\,\ref{fig:LowEnergyDoS_Altland}(b).
\textcolor{black}{Since critical configurations ($u\!=\!u_{c}^{j})$
have zero statistical measure, FSR would seemingly imply a vanishing
average DoS at $\varepsilon\!=\!0$. In the remainder of this paper,
we show that the difficulty arising from the discontinuous $\delta_{j}\!\left(\!\varepsilon,\!u\!=\!u_{\text{c}}\!\right)$
can be overcome by carefully accounting for the level statistics in
the infinite volume limit of a DWSM with random impurities. Such a
procedure does not alter the fate of $\rho\left(\varepsilon\!=\!0\right)$
in the presence of a finite concentration of impurities with a fixed
$u$\,\citep{Buchhold2018_2,Nandkishore2014}. However, the conclusions
on the stability of a DWSM are changed dramatically if a }\textit{\textcolor{black}{continuous
statistical diversity of ``near-critical'' impurities}}\textcolor{black}{{}
exists. }

\vspace{-0.6cm}

\section{\label{sec:Near-critical-impurities-lift}Near-critical impurities
lift the nodal density of states}

\vspace{-0.2cm}

\noindent \textcolor{black}{A finite concentration of exactly critical
wells or plateaus would introduce a macroscopic number of nodal bound
states. However, for a diversity of random impurities with potential
strengths drawn from a probability distribution function $p(u)$,
such fine-tuned configurations appear with zero probability and cannot
yield statistically significant contributions to the bulk nodal DoS\,\citep{Buchhold2018,Buchhold2018_2}.
Nevertheless, we find that low-energy resonances due to }\textit{\textcolor{black}{near-critical}}\textcolor{black}{{}
configurations ($u\!\approx\!u_{c}^{j}$) provide such a contribution.~The
phase shifts of such impurities signal the emergence of the zero-energy
bound states by a sharp resonance, namely, a quick $\pi$ variation
of $\delta_{j}\!\left(\varepsilon\right)$ as $u\!\to\!u_{c}^{j}$
originated in the valence band, which moves towards $\varepsilon\!=\!0$
and becomes sharper while always keeping $\delta_{j}\!\left(0,u\right)\!=\!0$.\,At
$u\!=\!u_{c}$, the situation is delicate because $\delta_{j}\!\left(\varepsilon\right)$
is no longer differentiable at $\varepsilon\!=\!0$.\,In that case,
one must work with Eq.\,\eqref{eq:Number of States Variation-2}
directly, and since there is a zero-energy $\pi$ discontinuity in
the phase shifts, a Dirac-$\delta$ distribution around the $u_{\text{c},n}^{j}$
emerges as the limit of plateau functions with a conserved integral
equal to $4$\,(the degeneracy of the $j\!=\!1/2$ single-impurity
bound states).\,This limit is depicted in Fig.\,\ref{fig:LowEnergyDoS_Altland}(a).\,An
immediate implication is that a DWSM is unstable to dilute random
impurities provided $p(u_{c}^{j})\!\neq\!0$ for at least one critical
$u_{c}^{j}$.\,Such a condition implies that the stability of a 3D
semimetallic phase ultimately depends on the type of impurity model
and the disorder statistics, i.e., whether it supports the resonant
mechanism driven by a continuous distribution of near-critical impurities.
These findings, supported below by accurate lattice simulations, show
that dirty 3D DWSMs with near-critical impurities are inherently unstable,
which sheds light on the previously reported AQC\,\citep{Pixley2016_2,pixley2016_3,Syzranov_2018,Ziegler2018,Goncalves2020,Wilson2020}.}

\textcolor{black}{In Fig.\,\ref{fig:LowEnergyDoS_Altland}(d), we
plot the change in the $j\!\!=\!\!1/2$ DoS due to a dilute diversity
of}\textit{\textcolor{black}{{} ``near-critical impurities}}\textcolor{black}{''.
The diversity is characterized by a Gaussian distribution $p(u)\!=\!\exp\left[-\left(u\!-\!\pi\right)/2\sigma^{2}\right]/\sqrt{2\pi}\sigma$
around $u_{\text{c}}\!=\!\pi$. The DoS is clearly lifted around $\varepsilon\!=\!0$,
forming a sharp symmetrical bump. For this diversity model, the peak
is Gaussian shaped near its center, and the corresponding area is
conserved as $\sigma\!\to\!0$. In this limit, a $4\delta\!\left(\varepsilon\right)$
distribution forms, i.e., all impurities are critical, each having
a fourfold degenerate zero-energy bound state.}

\vspace{-0.5cm}

\section{\label{sec:Lattice-simulations}Lattice simulations}

\vspace{-0.2cm}

Our prediction for the lifting of the DoS due to near-critical impurities
has been based on a continuum model for a single-node Dirac semimetal.
However, real Dirac materials and numerical simulations live in the
realm of lattice models, featuring several nodes and warped band structures.
To validate our previous conclusions, we perform real-space simulations
on a simple cubic lattice ($\mathcal{L}_{C}$ of parameter $a$ and
linear size $L$) with a four-orbital Hamiltonian derived from the
continuum Hamiltonian $\mathcal{H}$\,\citep{Pixley2016,Pixley2016_2,pixley2016_3},
namely,

\begin{figure}[t]
\begin{centering}
\hspace{-0.3cm}\includegraphics[scale=0.24]{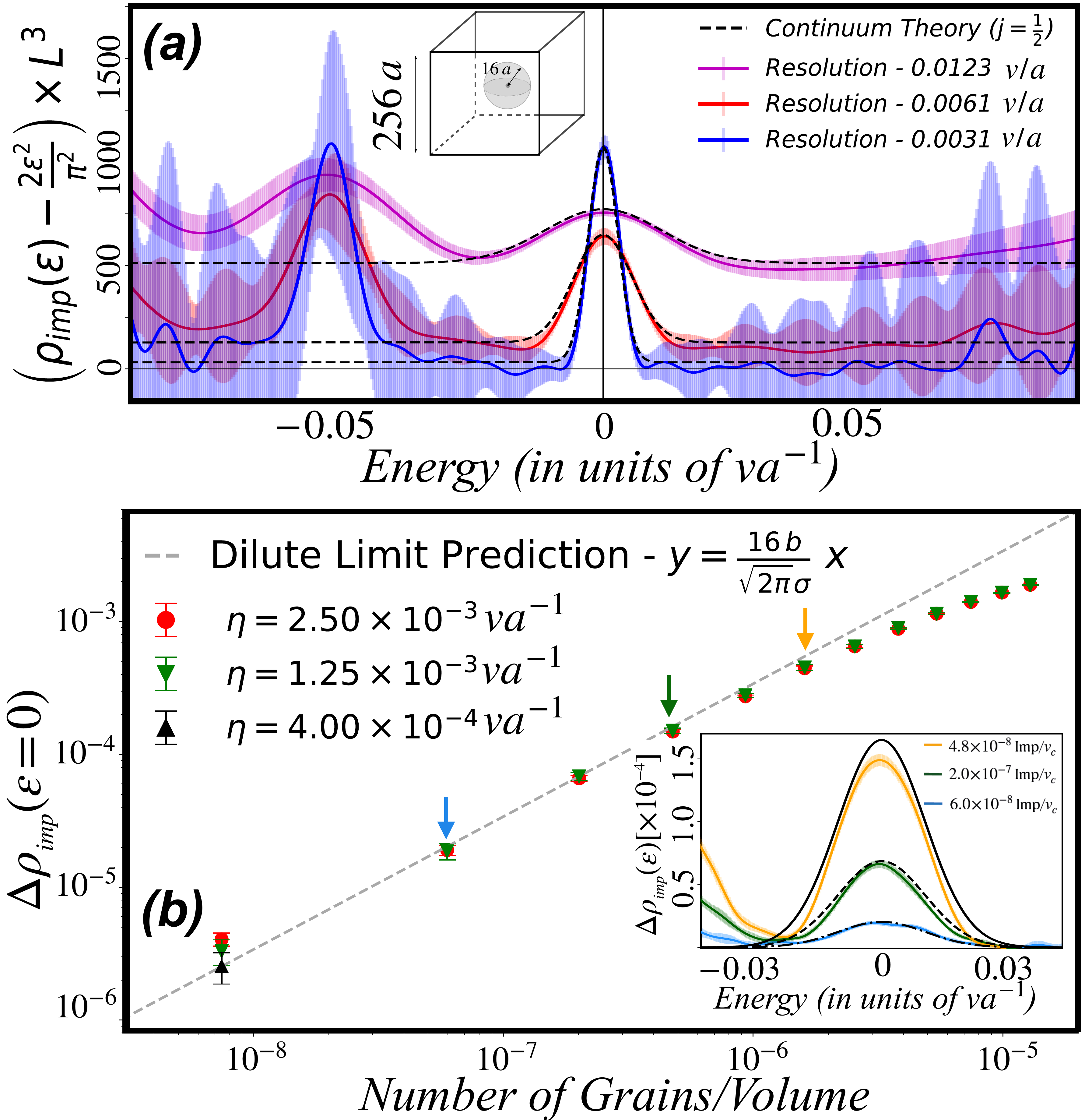}
\par\end{centering}
\vspace{-0.3cm}

\caption{\label{fig:LowEnergyDoS}(a) DoS change due to an impurity of critical
strength $u\!=\!\pi\,v/a$ and radius $16\,a$ inside a supercell
of $256^{3}$ sites. Vertical widths are $95\%$ statistical error
bars, and dashed lines are the continuum theory predictions. (b) Plot
of $\protect\av{\rho\!\left(E\!=\!0\right)}_{u}$ with several impurities
of radius $16\,a$ inside the simulated supercell of $512^{3}$ sites
for different resolutions $\eta$. The gray line is the dilute regime
prediction. The inset shows converged $\protect\av{\rho\!\left(E\right)}_{u}$
for three concentrations against the predictions of Fig.\,\ref{fig:LowEnergyDoS_Altland}(d)
(black lines).\vspace{-0.7cm}
}
\end{figure}


\vspace{-0.5cm}

\begin{align}
\negthickspace\!H & =\!\negthickspace\sum_{{\scriptscriptstyle \mathbf{R}\in\mathcal{L_{\text{C}}}}}\negthickspace\left[\frac{\imath v}{2a}\Psi_{\mathbf{{\scriptscriptstyle R}}}^{\dagger}\!\cdot\!\alpha^{j}\!\cdot\!\Psi_{{\scriptscriptstyle \mathbf{R}+a\hat{e}_{j}}}\!+\!\frac{\mathcal{U}\left(\mathbf{R}\right)}{2}\Psi_{\mathbf{{\scriptscriptstyle R}}}^{\dagger}\!\cdot\!\Psi_{\mathbf{{\scriptscriptstyle R}}}\!+\!\text{H.c.}\right]\!.\label{eq:LatticeModel}
\end{align}

\vspace{-0.2cm}

\noindent The DoS is calculated by means of accurate Chebyshev polynomial
expansions of the resolvent operator~$\delta(E\!-\!H)$\,\footnote{\textcolor{black}{In all numerical calculations, the variable $E$
is to be understood as a dimensionless energy, measured in units of
$a/\hbar v$.}} in very large systems as implemented in the open-source quantum transport
code $\mathtt{QUANTUM}\;\mathtt{KITE}$\,\citep{Joao2020}. The energy
resolution reads as $\eta\!=\!\pi\Delta E/2M$, where $\Delta E$
is the bandwidth of the Hamiltonian matrix and $M$ is the truncation
order of the polynomial expansion\,\citep{Weise2006,Ferreira2015}.
The DoS is obtained from $\rho(E)\!=\!\lim_{\eta\rightarrow0}\lim_{D\rightarrow\infty}D^{-1}\textrm{Tr}[\langle\delta_{\eta}(E\!-\!H)\rangle]$,
where $\langle...\rangle$ denotes disorder averaging and $D\!=\!4L^{3}$
is the Hilbert space dimension. \textcolor{black}{In order to simulate
systems with a vanishing mean-level spacing, thereby performing calculations
bounded only by $\eta$, we randomly sample over twisted boundaries\,\citep{Goncalves2020,Wilson2020}.
}This approach allowed the DoS to be calculated with unprecedented
working spectral resolutions as low as $\eta\!\simeq\!4\!\times\!10^{-4}v/a$\textcolor{black}{,
whose full convergence requires $M\!\approx\!30\,000$ polynomials.
}More technical details are provided in Appendix\,\ref{sec:AppendixD}.

\textcolor{black}{Figure\,\ref{fig:LowEnergyDoS}(a) shows the average
DoS induced by critical impurities, $\Delta\rho\left(E\right)\!=\!\rho_{\text{imp}}\left(E,u\!=\!\pi\right)\!-\!2E^{2}/\pi^{2}$,
in the dilute regime. The numerical data are compared with our analytical
results {[}Eq.\,\eqref{eq:Number of States Variation-2} and discussion
thereafter{]}, including the eightfold valley degeneracy, and properly
convoluted with Gaussian functions of width $\eta$ to mimic the finite
numerical spectral resolution. The lifting of the DoS at the node
and the underlying near-critical impurity mechanism are borne out
by the spectral calculations, which show excellent quantitative agreement
with the continuum theory, provided the impurity radius is large enough
{[}see Fig.\,\ref{fig:LowEnergyDoS}(a) and additional numerical
evidence in Appendix\,\ref{sec:AppendixD}{]}. In Fig.\,\ref{fig:LowEnergyDoS}(b),
we present an analogous calculation for a system having several impurities
inside the supercell. The impurities are placed randomly without superpositions,
and their strengths drawn from a Gaussian distribution $\text{\ensuremath{\mathcal{N}}}\!\!\left(\mu\!=\!\pi\,v/a\!\!,\sigma\!=\!0.3v/b\right)$.
The continuum theory prediction for the low-energy bump in the DoS
is reproduced in the diluted limit {[}see inset of Fig.\,\ref{fig:LowEnergyDoS}(b){]},
and the law $\rho\!\left(E\!=\!0\right)\!\propto\!c$ remains accurate
up to $10^{-6}$ impurities per unit cell. The overshooting for higher
concentrations is due to multi-impurity effects, which become more
effective as impurities are pushed closer together. }

\vspace{-0.3cm}

\section{\label{sec:Outlook}Conclusions and Outlook}

\vspace{-0.2cm}

\textcolor{black}{Here, we have shown that AQC must occur in 3D Dirac
semimetals having dilute short-range scalar impurities, if their random
parameters have a nonzero probability density at so-called }\textit{\textcolor{black}{magical}}\textcolor{black}{{}
values, where nodal bound states appear. These results were based
on a continuum formulation of the problem treated at the single-impurity
level and quantitatively confirmed by high-resolution lattice simulations
in a gapless }\textcolor{black}{\emph{multivalley}}\textcolor{black}{{}
Dirac model hosting $\approx10^{-9}\!\!\!\!-\!10^{-6}$ random scalar
impurities impurities per unit cell. The perfect agreement between
theory and numerical simulations gives confidence that the newly unveiled
resonant mechanism stemming from diverse near-critical impurities
is a crucial piece in the DWSM quantum criticality puzzle. Moreover,
disparities with previous work\,\citep{Buchhold2018,Buchhold2018_2}
are explained by the presence of physical $\pi$ jumps in the scattering
phase shifts that prevent a direct use of FSR for diverse impurities
around the aforementioned }\textit{\textcolor{black}{magical}}\textcolor{black}{{}
parameters. Similar conclusions are expected to hold for Weyl semimetals,
as scalar impurities do not couple the different Weyl sectors in the
infinite volume limit of our Dirac model. Meanwhile, our lattice nodal
DoS calculations show a crossover from a dilute regime at very low
impurity concentration (with DoS scaling linearly with $c$) to an
intermediate impurity density regime ($c\gtrsim10^{-6}$ impurities
per unit cell), where the DoS diverges from the analytical prediction.
This behavior can be traced to quantum-coherent multiple-impurity
scattering events, which are neglected in our continuum theory.}

\textcolor{black}{A related, but nontrivial, question concerns the
validity of these conclusions when dealing with lattice models having
uncorrelated on-site disorder. In light of our theory, as well as
earlier work\,\citep{Pixley2016_2,Ziegler2018}, one reasonably expects
the semimetallic phase to be unstable for unbounded distributions.
However, such systems with highly concentrated and atomic-sized (on-site)
impurities are exactly in the regime where the lattice results deviate
from continuum predictions, hinting at yet further subtleties when
relating the fate of disordered DWSM phases with rare-event bound
states.}

\vspace{-0.3cm}

\section{\textup{Acknowledgments}}

\vspace{-0.2cm}

We acknowledge support by the Portuguese Foundation for Science and
Technology through Strategic Funding No. UIDB/04650/2020, Projects
No. POCI-01-0145-FEDER-028887 (J.P.S.P. and J.M.V.P.L) and No. CEECIND/02936/2017
(B.A.), and Grant No. PD/BD/142774/2018 (J.P.S.P.). A.F. acknowledges
financial support from the Royal Society through a Royal Society University
Research Fellowship. The numerical calculations were performed on
the Viking Cluster, a high performance computing\,(HPC) facility
provided by the University of York. We are grateful for support from
the University of York's HPC service and their team. The authors are
grateful to A. Altland, M. Buchhold, M. Gonçalves, S.\,M. João, A.
Antunes, and J.\,M.\,B. Lopes dos Santos for valuable discussions.
The authors finally thank the anonymous referees for useful comments
which helped to improve the clarity of this work.

\vspace{-0.25cm}

\appendix

\section{\textcolor{black}{\label{sec:AppendixA}Spherical Dirac States}}

\vspace{-0.2cm}

\noindent Here, we provide technical details on the calculations leading
from the Dirac eigenvalue problem in the presence of a single central
scalar potential: $\mathcal{H}\!=\!\mathcal{H}_{0}\!+\!\mathcal{U}\!\left(\abs{\mathbf{r}}\right)$.
These results form the theoretical foundation for the main results
presented in this paper. Since the methods employed are scattered
around the existing literature\,\citep{Ma85,Ma_2006,Nandkishore2014},
we provide a detailed description of procedures to make the presentation
self-contained.

\vspace{-0.25cm}

\subsubsection{Derivation of the Radial Dirac Equations and Radial Eigenstates}

\vspace{-0.2cm}

\noindent The eigenvalue problem for an independent Dirac particle
in the presence of a central potential corresponds to finding the
solutions of

\vspace{-0.45cm}

\begin{align}
\mathcal{H}_{\mu\nu}\!\Psi_{\nu}\!\left(\mathbf{r}\right)\! & =\!\left[-\iota v_{\text{F}}\boldsymbol{\alpha}_{\mu\nu}\!\cdot\!\boldsymbol{\nabla}\!\!+\!\mathcal{U}\left(\abs{\mathbf{r}}\right)\delta_{\mu\nu}\right]\Psi_{\nu}\!\left(\mathbf{r}\right)\label{eq:DiracHamiltonian-1}\\
 & \qquad\qquad\qquad\qquad\qquad=\!\delta_{\mu\nu}E\Psi_{\nu}\!\left(\mathbf{r}\right),\nonumber 
\end{align}

\vspace{-0.2cm}

\noindent where the repeated Greek indices are summed over the four-spinor
components of the single-particle Dirac wavefunction. In particular,
we are interested in the special case of a potential well or plateau,
such that $\mathcal{U}\left(\abs{\mathbf{r}}\!<\!b\right)\!=\!\lambda$
and $\mathcal{U}\left(\abs{\mathbf{r}}\!\geq\!b\right)\!=\!\lambda$.

The first technical step towards solving Eq.\,(\ref{eq:DiracHamiltonian-1})
is to use a spherical coordinate system, $\left(r,\theta,\varphi\right)$,
and achieve a separation of variables. The way to do this is well
known in the relativistic quantum mechanics literature and is based
on identifying the orbital and spin angular momentum operators for
this system, which read

\vspace{-0.45cm}

\begin{equation}
\mathbf{L}\!=\!\iota\mathbb{I}_{4\times4}\varepsilon^{ijk}x_{j}\frac{\partial}{\partial x_{k}}\text{ and }\mathbf{S}\!=\!\frac{1}{2}\left(\begin{array}{cc}
\sigma^{i} & \mathbb{O}_{2\times2}\\
\mathbb{O}_{2\times2} & \sigma^{i}
\end{array}\right),\label{eq:OrbitalAngularMomentum}
\end{equation}
where the matrices act in the Dirac spinor indices. These quantities
are not conserved by the Hamiltonian $\mathcal{H}$; however, we can
build three mutually commuting observables out of $\mathbf{L}$ and
$\mathbf{S}$, which are conserved and uniquely define the spinor
and angular structure of the eigenfunctions of $\mathcal{H}$. These
are

\vspace{-0.5cm}

\begin{subequations}
\begin{equation}
J_{z}=\left(\begin{array}{cc}
L_{z}+\frac{1}{2}\sigma^{z} & \mathbb{O}_{2\times2}\\
\mathbb{O}_{2\times2} & L_{z}+\frac{1}{2}\sigma^{z}
\end{array}\right),\label{eq:Jz_Matrix4x4}
\end{equation}

\vspace{-0.3cm}

\begin{equation}
\abs{\mathbf{J}}^{2}\!\!=\!\abs{\mathbf{L}\!+\!\mathbf{S}}^{2}\!\!=\!\left(\!\!\!\begin{array}{cc}
\abs{\mathbf{L}}^{2}\!+\!\frac{3}{4}\!+\!\sigma^{i}L_{i} & \mathbb{O}_{2\times2}\\
\mathbb{O}_{2\times2} & \abs{\mathbf{L}}^{2}\!+\!\frac{3}{4}\!+\!\sigma^{i}L_{i}
\end{array}\!\!\right)\label{eq:Jz_Matrix4x4-1}
\end{equation}

\vspace{0.1cm}

\noindent and also $\mathcal{K}\!=\!\gamma^{0}\!\cdot\!(2L_{i}S_{i}\!-\!1)$,
which explicitly reads

\vspace{-0.5cm}

\begin{equation}
\mathcal{K}=\left(\begin{array}{cc}
\sigma^{i}L_{i}+\mathbb{I}_{2\times2} & \mathbb{O}_{2\times2}\\
\mathbb{O}_{2\times2} & -\sigma^{i}L_{i}-\mathbb{I}_{2\times2}
\end{array}\right).\label{eq:K_Matrix4x4}
\end{equation}
\end{subequations}
 It is easy to verify that all three operators in Eqs.\,\eqref{eq:Jz_Matrix4x4}---\eqref{eq:K_Matrix4x4}
commute among themselves and with $\mathcal{H}$. Crucial for the
latter is the fact that $\mathcal{U}\!\left(\mathbf{r}\right)\!=\!\mathcal{U}\!\left(r\right)$,
which guarantees that the impurity does not break rotational symmetry
around its center. Therefore a common eigenbasis of $\mathbf{\abs J^{2}},$
$J_{z}$,, and $\mathcal{K}$ can be built and labeled by the set
of quantum numbers $j\!\in\!\left\{ \nicefrac{1}{2},\nicefrac{3}{2},\cdots\right\} $,
$j_{z}\!\in\!\left\{ -j,-j+1,\cdots,j\right\} $, and $\kappa\!=\!\pm1$.
The quantum number $\kappa$ appears by solving $\mathcal{K}$'s eigenvalue
problem,

\vspace{-0.5cm}

\begin{equation}
\mathcal{K}\Psi_{j,j_{z}}\left(\mathbf{r}\right)=\hbar^{2}\kappa\left(j+\frac{1}{2}\right)\Psi_{j,j_{z}}\left(\mathbf{r}\right).
\end{equation}
Using the previous operators, a general form for the eigenspinors
indexed by the set $\left(j,j_{z},\kappa\right)$ is

\vspace{-0.35cm}

\begin{equation}
\Psi_{j,j_{z}}^{\kappa}\left(r,\theta,\varphi\right)=\frac{1}{r}\left(\begin{array}{c}
f_{j}^{\kappa}\left(r\right)\Theta_{j,j_{z}}^{-\kappa}\left(\theta,\varphi\right)\\
\iota g_{j}^{\kappa}\left(r\right)\Theta_{j,j_{z}}^{\kappa}\left(\theta,\varphi\right)
\end{array}\right),\label{eq:Form_Spinors}
\end{equation}
\vspace{-0.2cm}

\noindent where $f_{j}^{\kappa}\!\left(r\right)/g_{j}^{\kappa}\!\left(r\right)$
are radial functions and $\Theta^{\kappa}\!\left(\theta,\varphi\right)$
are spin-$\nicefrac{1}{2}$ spherical harmonics

\begin{subequations}
\vspace{-0.4cm}

\begin{equation}
\Theta_{j,j_{z}}^{+}\left(\theta,\varphi\right)=\left(\begin{array}{c}
\sqrt{\frac{j-j_{z}+1}{2j+2}}Y_{j_{z}-1/2}^{j+1/2}\left(\theta,\varphi\right)\\
-\sqrt{\frac{j+j_{z}+1}{2j+2}}Y_{j_{z}+1/2}^{j+1/2}\left(\theta,\varphi\right)
\end{array}\right)\label{eq:Normalized+-3-1}
\end{equation}
and

\vspace{-0.7cm}

\begin{equation}
\Theta_{j,j_{z}}^{-}\left(\theta,\varphi\right)=\left(\begin{array}{c}
\sqrt{\frac{j+j_{z}}{2j}}Y_{j_{z}-1/2}^{j-1/2}\left(\theta,\varphi\right)\\
\sqrt{\frac{j-j_{z}}{2j}}Y_{j_{z}+1/2}^{j-1/2}\left(\theta,\varphi\right)
\end{array}\right)\label{eq:Normalized+-3-1-1}
\end{equation}
\end{subequations}
 which in this form are orthonormalized in the unit sphere, i.e.,

\vspace{-0.25cm}

\begin{equation}
\!\!\!\int_{0}^{\pi}\!\!\!\!\!\sin\!\theta d\theta\!\!\!\int_{0}^{2\pi}\!\!\!\!\!\!\!\!d\varphi\!\left[\Theta_{j\!,j_{z}}^{\kappa}\!\!\left(\theta,\varphi\right)\right]^{\dagger}\!\!\!\cdot\Theta_{j'\!,j_{z}^{'}}^{\kappa'}\!\!\left(\theta,\varphi\right)\!=\!\delta_{j,j'}\!\delta_{j_{z},j_{z}^{'}}\!\delta_{\kappa,\kappa'}.\label{eq:OrthogonalitySpinSphericalHarmonics}
\end{equation}
Besides the orthonormality condition of Eq.\,\eqref{eq:OrthogonalitySpinSphericalHarmonics},
$\Theta_{j,j_{z}}^{\kappa}\!\!\left(\Omega\right)$ have some further
useful properties, namely,

\vspace{-0.3cm}

\noindent 
\begin{subequations}
\begin{equation}
\boldsymbol{\sigma}\!\cdot\!\hat{\mathbf{r}}\,\Theta^{\kappa}\!\left(\theta,\varphi\right)\!=\!\left(\boldsymbol{\sigma}\!\cdot\!\hat{\mathbf{r}}\right)^{2}\Theta^{\kappa}\!\left(\theta,\varphi\right)=\Theta^{-\kappa}\!\left(\theta,\varphi\right)\label{eq:MixingTheta}
\end{equation}

\vspace{-0.55cm}

\begin{equation}
\boldsymbol{\sigma}\!\cdot\!\mathbf{L}\Theta^{+}\!\left(\theta,\varphi\right)\!=\!-\hbar\left(j\!+\!\frac{3}{2}\right)\Theta^{+}\!\left(\theta,\varphi\right)\label{eq:J_zeigenvalues-1-1}
\end{equation}

\vspace{-0.55cm}

\begin{equation}
\boldsymbol{\sigma}\!\cdot\!\mathbf{L}\,\Theta^{-}\!\left(\theta,\varphi\right)\!=\!\hbar\left(j\!-\!\frac{1}{2}\right)\Theta^{-}\!\left(\theta,\varphi\right),\label{eq:J_zeigenvalues-1-1-1}
\end{equation}
\end{subequations}
 where $\boldsymbol{\sigma}\!=\!\left(\sigma^{x},\sigma^{y},\sigma^{z}\right)$
is a vector of Pauli matrices and the scalar products are to be understood
as a summation over spacial indices. Finally, we can proceed and write
the Hamiltonian $\mathcal{H}$ explicitly as a differential operator
in spherical coordinates. That way, it reads

\vspace{-0.2cm}

\begin{equation}
\mathcal{H}\!=\!\!\left(\!\!\begin{array}{cc}
\mathcal{U}\!\left(r\right)\!\mathbb{I}_{2\times2} & -\iota v\boldsymbol{\sigma}\!\cdot\!\mathbf{\hat{r}}\left[\partial_{r}\!-\!\frac{\boldsymbol{\sigma}\cdot\mathbf{L}}{r}\right]\\
-\iota v\boldsymbol{\sigma}\!\cdot\!\mathbf{\hat{r}}\left[\partial_{r}\!-\!\frac{\boldsymbol{\sigma}\cdot\mathbf{L}}{r}\right] & \mathcal{U}\!\left(r\right)\!\mathbb{I}_{2\times2}
\end{array}\!\!\right).\label{eq:HamiltonianSpherical}
\end{equation}
Using this form for $\mathcal{H}$, one can plug spinors as in Eq.\,\eqref{eq:Form_Spinors}
into the eigenvalue problem of Eq.\,\eqref{eq:DiracHamiltonian-1}
and arrive at the following coupled systems of ordinary differential
equations:

\noindent \vspace{-0.6cm}

\begin{equation}
\!\!\!\!\!\begin{cases}
\!\frac{d}{dr}\!g_{j,E}^{\kappa}\!\left(r\right)\!\pm\!\frac{1}{r}\left(j\!+\!\frac{1}{2}\right)\!g_{j,E}^{\kappa}\!\left(r\right)\!=\!\!\frac{1}{v}\!\left[E\!-\!\mathcal{U}\!\left(r\right)\right]\!f_{j,E}^{\kappa}\!\left(r\right)\\
\!\frac{d}{dr}\!f_{j,E}^{\kappa}\left(r\right)\!\mp\!\frac{1}{r}\left(j\!+\!\frac{1}{2}\right)\!f_{j,E}^{\kappa}\!\left(r\right)\!=\!\!\frac{1}{v}\!\left[\mathcal{U}\!\left(r\right)\!-\!E\right]\!g_{j,E}^{\kappa}\!\left(r\right)
\end{cases}\!\!\!\!\!\!\!\!.\label{eq:Sys1-1}
\end{equation}
In the case of the spherical well or plateau that concerns us, Eq.\,(\ref{eq:Sys1-1})
reduces to either

\noindent \vspace{-0.6cm}

\begin{equation}
\begin{cases}
\!\frac{d}{dx}\!g_{j,\varepsilon}^{\kappa}\!\left(x\right)\!\pm\!\frac{1}{x}\left(j\!+\!\frac{1}{2}\right)g_{j,\varepsilon}^{\kappa}\!\left(x\right)\!=\!\left(\varepsilon\!-\!u\right)\!f_{j,\varepsilon}^{\kappa}\!\left(x\right)\\
\frac{d}{dx}\!f_{j,\varepsilon}^{\kappa}\!\left(x\right)\!\mp\!\frac{1}{x}\left(j\!+\!\frac{1}{2}\right)\!f_{j,\varepsilon}^{\kappa}\!\left(x\right)\!=\!\left(u\!-\!\varepsilon\right)\!g_{j,\varepsilon}^{\kappa}\!\left(x\right)
\end{cases}\!\!\!\!\!\!\!\!,\label{eq:SystemInside}
\end{equation}
inside the impurity, or

\vspace{-0.5cm}

\begin{equation}
\begin{cases}
\frac{d}{dx}g_{j,\varepsilon}^{\kappa}\left(x\right)\pm\frac{1}{x}\left(j+\frac{1}{2}\right)g_{j,\varepsilon}^{\kappa}\left(x\right)=\varepsilon f_{j,\varepsilon}^{\kappa}\left(x\right)\\
\frac{d}{dx}f_{j,\varepsilon}^{\kappa}\left(x\right)\mp\frac{1}{x}\left(j+\frac{1}{2}\right)f_{j,\varepsilon}^{\kappa}\left(x\right)=-\varepsilon g_{j,\varepsilon}^{\kappa}\left(x\right)
\end{cases},\label{eq:SystemOutside}
\end{equation}

\noindent outside of it. In Eqs.\,\eqref{eq:SystemInside} and \eqref{eq:SystemOutside},
we use dimensionless scales, namely, $x\!=\!r/b$, $\varepsilon\!=\!Eb/v$,
and $u\!=\!\lambda b/v$. The solutions inside the impurity (as long
as $\varepsilon\!\neq\!u$) always have the general form

\vspace{-0.3cm}

\begin{subequations}
\begin{align}
g_{j,\varepsilon}^{+}\left(x\!<\!1\right) & =\!\mathcal{A}^{+}\!\sqrt{x}J_{j+1}\left(\abs{\varepsilon\!-\!u}\!x\right),\label{eq:g+FreeSolution-3-2}\\
f_{j,\varepsilon}^{+}\left(x\!<\!1\right) & =\!\mathcal{A}^{+}\!\text{sgn}\left(\varepsilon\!-\!u\right)\sqrt{x}J_{j}\left(\abs{\varepsilon\!-\!u}\!x\right),\\
g_{j,\varepsilon}^{-}\left(x\!<\!1\right) & =\!\mathcal{A}^{-}\!\text{sgn}\left(\varepsilon\!-\!u\right)\sqrt{x}J_{j}\left(\abs{\varepsilon\!-\!u}\!x\right),\\
f_{j,\varepsilon}^{-}\left(x\!<\!1\right) & =\!\mathcal{A}^{-}\!\sqrt{x}J_{j+1}\!\left(\abs{\varepsilon\!-\!u}\!x\right),\label{eq:f-FreeSolution}
\end{align}
\end{subequations}

\noindent where $\mathcal{A}^{\pm}$ are complex adjustable constants.
Outside the impurity and for non-zero energy, one has instead

\vspace{-0.3cm}

\noindent 
\begin{subequations}
\begin{align}
g_{j,\varepsilon}^{+}\left(x\!>\!1\right)\!= & \mathcal{B}^{+}\!\sqrt{x}\left[\cos\delta_{j}^{+}\!\left(\varepsilon\right)J_{j+1}\left(\abs{\varepsilon}\!x\right)\right.\label{eq:g+FreeSolution-3-1-1-1}\\
 & \qquad\qquad\left.-\text{sgn}\!\left(\varepsilon\right)\sin\delta_{j}^{+}\!\left(\varepsilon\right)Y_{j+1}\left(\abs{\varepsilon}\!x\right)\right]\nonumber \\
f_{j,\varepsilon}^{+}\left(x\!>\!1\right)\!= & \mathcal{B}^{+}\!\sqrt{x}\left[\cos\delta_{j}^{+}\left(\varepsilon\right)J_{j+1}\left(\abs{\varepsilon}\!x\right)\right.\\
 & \qquad\qquad\left.-\sin\!\delta_{j}^{+}\left(\varepsilon\right)Y_{j}\left(\abs{\varepsilon}\!x\right)\right]\nonumber \\
f_{j,\varepsilon}^{-}\left(x\!>\!1\right)\!= & \mathcal{B}^{-}\!\sqrt{x}\left[\cos\delta_{j}^{-}\left(\varepsilon\right)J_{j+1}\left(\abs{\varepsilon}\!x\right)\right.\\
 & \qquad\qquad\left.-\text{sgn}(\varepsilon)\sin\!\delta_{j}^{-}\left(\varepsilon\right)Y_{j+1}\left(\abs{\varepsilon}\!x\right)\right]\nonumber \\
g_{j,\varepsilon}^{-}\left(x\!>\!1\right)\!= & \mathcal{B}^{-}\!\sqrt{x}\left[\text{sgn}\left(\varepsilon\right)\cos\!\delta_{j}^{-}\!\left(\varepsilon\right)J_{j}\!\left(\abs{\varepsilon}\!x\right)\right.\label{eq:g-FreeSolutionOutside}\\
 & \qquad\qquad\left.-\sin\!\delta_{j}^{-}\left(\varepsilon\right)Y_{j}\left(\abs{\varepsilon}\!x\right)\right]\nonumber 
\end{align}
\end{subequations}

\noindent where the choice of parametrization in the linear combination
was made for convenience. Note that the exterior solutions feature
both $J_{n}$ and $Y_{n}$ components, being always regular and physically
admissible in their support ($x\!\geq\!1$). Now, all we must do is
constrain the functions $\delta_{j}^{\pm}\!\left(\varepsilon\right)$
such that the spinor $\Psi\!\left(\mathbf{r}\right)$ is continuous
at the impurity's surface\,($x\!=\!1$). Using Eqs.\,\eqref{eq:g+FreeSolution-3-1-1-1}---\eqref{eq:g-FreeSolutionOutside},
this implies that

\vspace{-0.4cm}

\begin{align}
\tan\!\delta_{j}^{\pm}\!\left(\varepsilon,u\right)\!\! & =\!\left[\text{sgn}\!\left(\varepsilon\!-\!u\right)\!J_{j+1}\!\left(\abs{\varepsilon}\right)\!J_{j}\!\left(\abs{\varepsilon\!-\!u}\right)\!\right.\label{eq:B/A-1-1-1-1}\\
 & \left.-\text{sgn}\!\left(\varepsilon\right)J_{j}\!\left(\abs{\varepsilon}\right)J_{j+1}\!\left(\abs{\varepsilon\!-\!u}\right)\right]/\left[\text{sgn}\left(\varepsilon\right)\right.\nonumber \\
 & \qquad\qquad\quad\times\text{sgn}\left(\varepsilon\!-\!u\right)Y_{j+1}\left(\abs{\varepsilon}\right)J_{j}\left(\abs{\varepsilon\!-\!u}\right)\nonumber \\
 & \qquad\qquad\qquad\qquad\qquad\left.-\!Y_{j}\left(\abs{\varepsilon}\right)J_{j+1}\!\left(\abs{\varepsilon\!-\!u}\right)\right].\nonumber 
\end{align}
This equation is independent of $\kappa$, which allows us to define
a unique function, $\delta_{j}\!\left(\varepsilon,u\right)$, for
both the $\kappa\!=\!\pm$ sectors, which appears a single twofold
degeneracy of the states in the problem. The previous facts justify
Eq.\,\eqref{eq:B/A-1-1-1} of Sec.\,\ref{sec:Continuum-theory}.

The previous analysis is valid for the entire spectrum, except at
the important $\varepsilon\!=\!0$ point. Here, the interior solutions
are the same, but the radial system outside the impurity decouples,
i.e.,

\vspace{-0.2cm}

\begin{equation}
\begin{cases}
\frac{d}{dx}g_{j,\varepsilon}^{\kappa}\left(x\right)\pm\frac{1}{x}\left(j+\frac{1}{2}\right)g_{j,\varepsilon}^{\kappa}\left(x\right)=0\\
\frac{d}{dx}f_{j,\varepsilon}^{\kappa}\left(x\right)\mp\frac{1}{x}\left(j+\frac{1}{2}\right)f_{j,\varepsilon}^{\kappa}\left(x\right)=0
\end{cases}.\label{eq:SystemOutside-1}
\end{equation}
The latter allows for power-law solutions, of which the physically
admissible ones (i.e., the ones decaying with $x$) are of the form

\vspace{-0.6cm}

\begin{subequations}
\begin{equation}
g_{j,\varepsilon=0}^{+}\!\left(x\!\geq\!1\right)\!=\!\frac{\mathcal{B}^{+}}{x^{j+1/2}}\;\text{and}\;f_{j,\varepsilon=0}^{+}\!\left(x\!\geq\!1\right)\!=\!0
\end{equation}

\vspace{-0.5cm}

\begin{equation}
g_{j,\varepsilon=0}^{-}\!\left(x\!\geq\!1\right)\!=\!0\;\text{and}\;f_{j,\varepsilon=0}^{-}\!\left(x\!\geq\!1\right)\!=\!\frac{\mathcal{B}^{-}}{x^{j+1/2}}.
\end{equation}
\end{subequations}
 Both these solutions, being joined continuously to the interior solutions
of Eqs.\,\eqref{eq:g+FreeSolution-3-2}---\eqref{eq:f-FreeSolution}
require that $J_{j}\left(\abs u\right)\!=\!0$. This condition gives
rise to a discrete set of parameters $u$, for which these bound-state
solutions are allowed.

Finally, we remark that all the eigenstates determined here (the unbound
and bound ones) have an intrinsic $2j+1$ degeneracy due to the rotational
invariance of the Hamiltonian. This degeneracy factor is explicitly
taken into account in all calculations done in the main text.

\section{\label{sec:AppendixB}Self-Adjoint Restriction of the Dirac Hamiltonian
to a Finite Sphere}

To derive the relation between the scattering phase shifts and the
change in the DoS due to a dilute diversity of impurities, we make
explicit use of the restriction of $\mathcal{H}$ to a finite sphere
of radius $R\gg b$, i.e., $S_{R}$. Restricting a continuum Hamiltonian
to a finite volume of space generally makes its action on the original
Hilbert space non-Hermitian. The way around this is to impose appropriate
boundary conditions which restrict the original basis to a subset,
generating a subspace inside of which the Hamiltonian preserves its
Hermiticity. This is called taking a self-adjoint extension of $\mathcal{H}$
to a finite domain.

In the case of the Dirac Hamiltonian with a scalar potential, $\mathcal{H}\!=\!-\iota v\boldsymbol{\alpha}\!\cdot\!\boldsymbol{\nabla}\!+\!\mathcal{U}\!\left(\mathbf{r}\right)$,
the Hermiticity condition is imposed by guaranteeing that for any
two Dirac spinor states $\Phi^{1}\left(\mathbf{r}\right)$ and $\Phi^{2}\left(\mathbf{r}\right)$,
the following condition holds:

\vspace{-0.4cm}

\begin{align}
\int_{S_{R}}\!\!\!\!d^{\text{3}}\mathbf{r} & \left[\Phi_{\mu}^{2}\left(\mathbf{r}\right)\right]^{\dagger}\left[-\iota v\boldsymbol{\alpha}_{\mu\nu}\!\cdot\!\boldsymbol{\nabla}\!+\!\mathcal{U}\!\left(\mathbf{r}\right)\delta_{\mu\nu}\right]\!\Phi_{\nu}^{1}\left(\mathbf{r}\right)\\
 & =\left[\int_{S_{R}}\!\!\!\!\!\!d^{\text{3}}\mathbf{r}\left[\Phi_{\mu}^{1}\left(\mathbf{r}\right)\right]^{\dagger}\!\left[-\iota v\boldsymbol{\alpha}_{\mu\nu}\!\cdot\!\boldsymbol{\nabla}\!+\!\mathcal{U}\!\left(\mathbf{r}\right)\delta_{\mu\nu}\right]\!\Phi_{\nu}^{2}\left(\mathbf{r}\right)\right]^{*}\!\!\!.\nonumber 
\end{align}

\noindent After some straightforward manipulation, this condition
can be cast into the equivalent form

\vspace{-0.4cm}

\begin{equation}
\varoiint_{\partial S_{R}}d^{\text{2}}S\left[\Phi_{\mu}^{1}\left(\mathbf{r}\right)\right]^{\dagger}\left[\boldsymbol{\alpha}_{\mu\nu}\!\cdot\!\hat{\mathbf{n}}\right]\Phi_{\nu}^{2}\left(\mathbf{r}\right)=0,\label{eq:NullCurrent}
\end{equation}
where $\hat{\mathbf{n}}\!=\!\left(n_{x},n_{y},n_{z}\right)$ is an
outwards unit vector normal to the spherical surface $\partial S_{R}$.
This is precisely the condition presented in Sec.\,\ref{sec:Impurity-induced-change-in}.
Unsurprisingly, Eq.\,\eqref{eq:NullCurrent} is easily interpreted
as guaranteeing that no net particle current crosses the boundary
of $S_{R}$, which expresses particle conservation implied by Hermiticity.

\begin{figure}[t]
\begin{centering}
\hspace{-0.2cm}\includegraphics[scale=0.15]{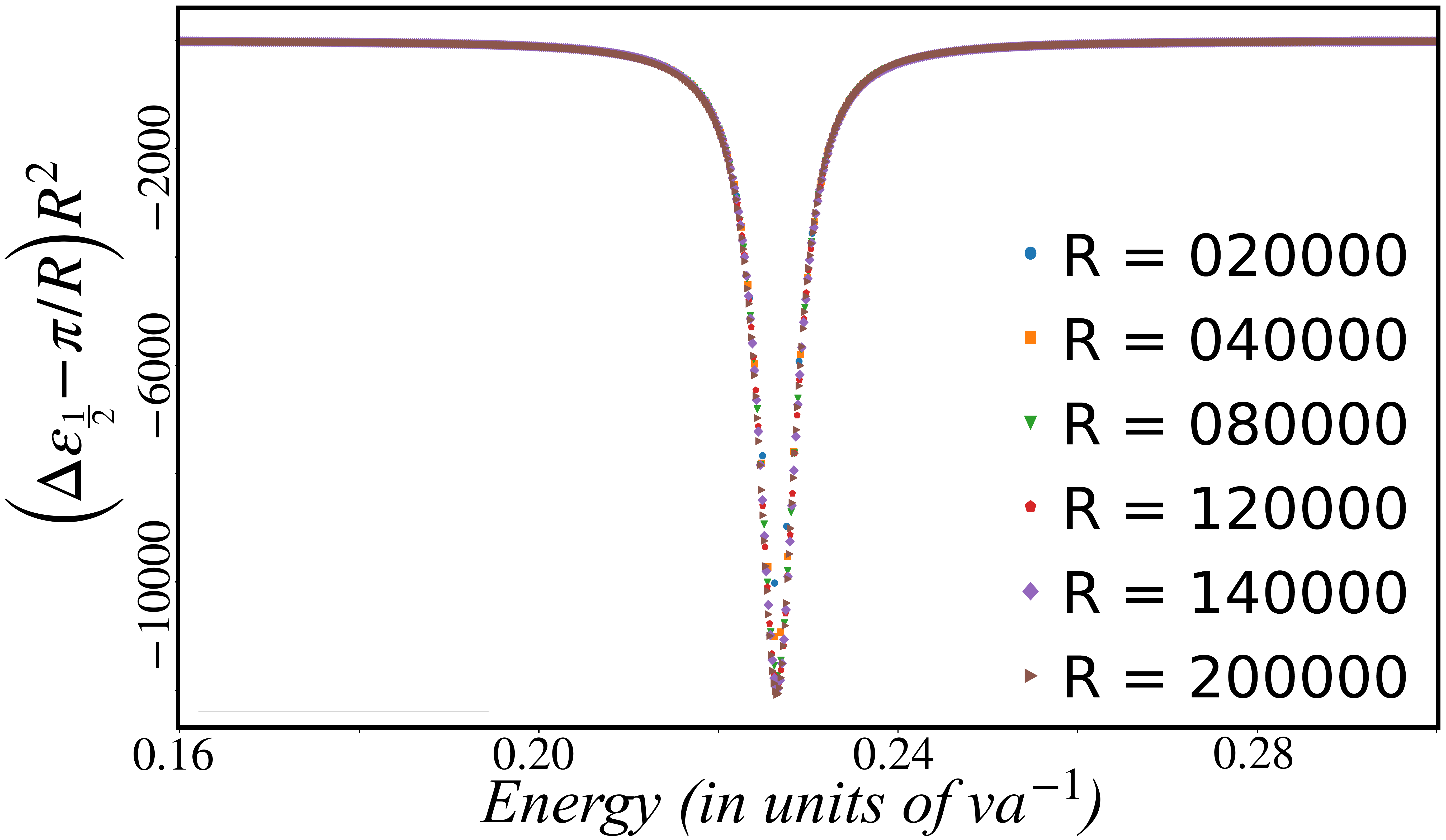} 
\par\end{centering}
\vspace{-0.3cm}

\caption{\label{fig:Spacings}Plot of $\Delta\varepsilon_{{\scriptscriptstyle 1/2}}\!=\!\varepsilon_{{\scriptscriptstyle n+1}}^{{\scriptscriptstyle 1/2}}\!-\!\varepsilon_{{\scriptscriptstyle n}}^{{\scriptscriptstyle 1/2}}$
the nearest-level spacings for a spherical impurity with $u\!=\!3.1867$
calculated from the numerically found solutions of the boundary condition
{[}Eq.\,(\ref{eq:BoundaryCondition}){]}. The data points are represented
as $R^{2}\!\times\!\left(\Delta\varepsilon_{{\scriptscriptstyle 1/2}}\!-\!\pi R^{-1}\right)$,
such that a collapse of different values of $R$ is achieved. This
collapse indicates that $\Delta\varepsilon_{{\scriptscriptstyle 1/2}}\left(R,\varepsilon,u\right)\!\approx\!\pi R^{-1}\!-\!f\left(\varepsilon,u\right)R^{-2}$
in the presence of an impurity. $R$ is measured in units of $b$.\vspace{-0.4cm}
 }
\end{figure}

Meanwhile, since all $\boldsymbol{\alpha}$ matrices are composed
of off-diagonal $2\!\times\!2$ blocks, one can easily see that Eq.\,\eqref{eq:NullCurrent}
is satisfied whenever we impose either the first or the last two components
of the Dirac spinors to be zero at $\partial S_{R}$. Considering
spinors of the form given in Eq.\,\eqref{eq:Form_Spinors}, such
a condition translates into either $f_{j}^{+}\!\left(R\right)\!=\!g_{j}^{-}\!\left(R\right)\!=\!0$
or $f_{j}^{-}\!\left(R\right)\!=\!g_{j}^{+}\!\left(R\right)\!=\!0$.
The other two combinations cannot be satisfied, as the zeros of Bessel
functions of different $j$'s never coincide. For the purposes of
this work, we chose the first of these conditions (although the specific
self-adjoint extension should not be relevant for any thermodynamic
limit results). Finally, by using the general form of the exterior
scattering solutions found earlier {[}Eqs.\,\eqref{eq:g+FreeSolution-3-1-1-1}---\eqref{eq:g-FreeSolutionOutside}{]},
we arrive at our final form for the boundary condition,

\vspace{-0.4cm}

\begin{equation}
\cos\!\delta_{j}\!\left(\varepsilon,u\right)J_{j}\!\left(\abs{\varepsilon}\!R\right)\!-\!\text{sgn}\left(\varepsilon\right)\sin\!\delta_{j}\!\left(\varepsilon\right)Y_{j}\left(\abs{\varepsilon}\!R\right)\!=\!0,\label{eq:BoundaryCondition}
\end{equation}

\noindent \vspace{-0.25cm}
 where $R$ is measured in units of $b$.

\subsubsection{Level Spacing of Central Impurity Dirac Hamiltonian}

First, we remark on an important consequence of the boundary condition
in Eq.\,\eqref{eq:BoundaryCondition}. This condition imposes a quantization
of energy levels, turning the continuous spectrum into a discrete
one with a density of levels that scales with $R$. Provided that
we are looking at finite energies ($\varepsilon\!\neq\!0$) and with
$\abs{\varepsilon}\!R\!\gg\!1$, Eq.\,\eqref{eq:BoundaryCondition}
can be taken in its asymptotic form, namely,

\vspace{-0.35cm}

\begin{equation}
\cos\left(\abs{\varepsilon}\!R\!+\!\frac{\pi}{2}\left(j\!+\!\frac{1}{2}\right)\!+\!\text{sgn}\left(\varepsilon\right)\delta_{j}\!\left(\varepsilon,u\right)\right)\!=\!0.\label{eq:HermiticityCondition-1-1}
\end{equation}
In the absence of an impurity, we have $\delta_{j}\!\left(\varepsilon,0\right)\!=\!0$,
and the mesh of energy levels allowed by the boundary conditions (in
a given $j$ sector) is simply $\varepsilon_{n}^{j}\!\approx\!\frac{n\pi}{R}\!+\!\text{sgn\ensuremath{\left(n\right)}}\!\frac{\pi}{2R}\left(j\!+\!1/2\right)$,
with $n\!\in\!\mathbb{Z}$. This yields a mean-level spacing which
is uniform across the spectrum and equal to $\pi R^{-1}$. In the
presence of the impurity (which induces energy-dependent phase shifts),
Eq.\,\eqref{eq:HermiticityCondition-1-1} does not seem to have a
simple solution. However, if $R$ is large enough such that $\delta_{j}\!\left(\varepsilon,u\right)$
is a slowly varying function across an energy interval of width $\pi R^{-1}$,
then one can say that the allowed energy levels are roughly

\vspace{-0.4cm}

\noindent 
\begin{equation}
\varepsilon_{n}^{j}\!\approx\!\frac{n\pi}{R}\!+\!\text{sgn\ensuremath{\left(n\right)}\!}\frac{\pi}{2R}\left(j\!+\!1/2\right)\!-\!\frac{\delta_{j}\left(\varepsilon_{n}^{j},u\right)}{R},
\end{equation}

\noindent \vspace{-0.1cm}
giving a correction to the mean-level spacing relative to the case
$u\!=\!0$, which is simply

\vspace{-0.4cm}

\begin{align}
\varepsilon_{n+1}^{j}\!-\!\varepsilon_{n}^{j}\! & \approx\!\frac{\pi}{R}\!-\!\frac{1}{R}\left[\delta_{j}\left(\varepsilon_{n}^{j}+\frac{\pi}{R},u\right)\!-\!\delta_{j}\!\left(\varepsilon_{n}^{j},u\right)\right]\\
 & \qquad\qquad\approx\!\frac{\pi}{R}\left[1\!-\!\frac{\pi}{R}\left.\frac{\partial}{\partial\varepsilon}\delta_{j}\left(\varepsilon,u\right)\right|_{\varepsilon=\varepsilon_{n}^{j}}\!\right],\nonumber 
\end{align}

\vspace{-0.3cm}

\noindent with an analogous expression for $n\!<\!0$. Hence, we conclude
that the correction to the mean-level spacing due to a single impurity
is always $\propto\!\mathcal{O}\!\left(R^{-2}\right)$, which is subleading
relative to the original $\pi/R$ spacing. This result is exemplified
by a numerical solution of Eq.\,\eqref{eq:BoundaryCondition} in
Fig.\,\ref{fig:Spacings} and justifies our arguments on the number
of states migrating in or out of an energy interval given in Sec.\,\ref{sec:Impurity-induced-change-in}.

\section{\label{sec:AppendixC}A Consistent Definition of the Scattering Phase
Shifts}

\noindent In this appendix, we use the spherical Dirac eigenstates
found earlier to define the scattering phase shifts in a way that
allows a direct relation to the impurity induced change in the density
of states. We recover early results which explain the crucial zero-energy
$\pi$-discontinuity observed in our calculations as due to the appearance
of critical bound states in the transition between the valence and
conduction band. This connects our results to \textit{Levinson's theorem}
applied to noninteracting Dirac particles.

The exterior scattering wave functions of the gapless Dirac equation
with a spherical well or plateau (of strength $\lambda$ and radius
$b$) were found to be of the form

$ $

\vspace{-0.5cm}

\noindent \onecolumngrid

\begin{figure*}
\vspace{-0.4cm}

\begin{centering}
\includegraphics[scale=0.26]{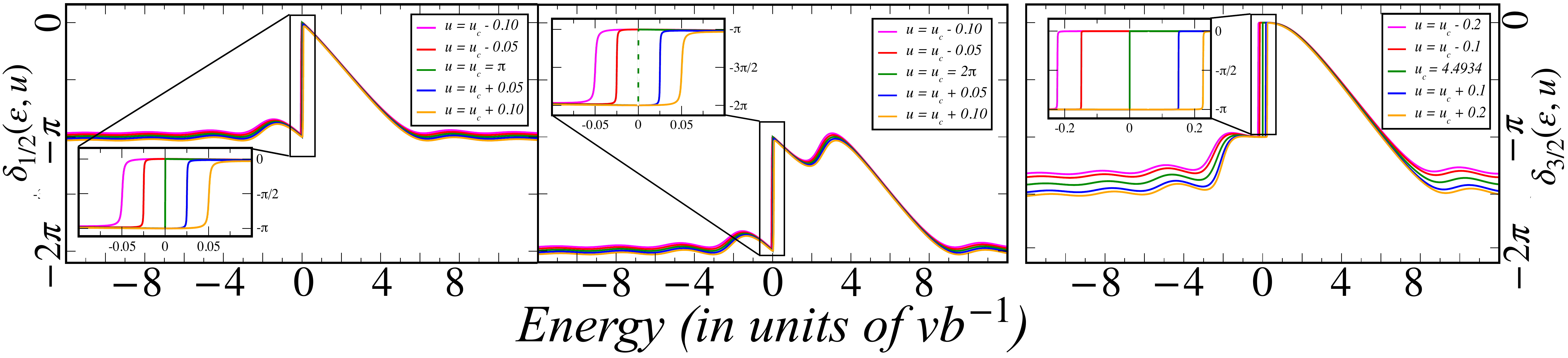} 
\par\end{centering}
\vspace{-0.4cm}

\caption{\label{fig:RadiusDependence_j1_2-1}Plots of the phase shifts for
$j\!=\!1/2$ (left and middle panels) around the two first critical
values $u\!=\!\pi,2\pi\hbar v_{\text{F}}a^{-1}$, and the first critical
value $u\!\approx\!4.4934\cdots$ for $j\!=\!3/2$ (right panel).
The main panels show the assigned asymptotic behavior, $\delta_{j}\left(\varepsilon\!\to\!\pm\infty,u\right)\!\to\!-\!u$,
in each case, while the insets depict the formation of a true $\pi$
discontinuity at $\varepsilon\!=\!0$ when $u\!=\!u_{\text{c}}^{j}$.\vspace{-0.5cm}
 }
\end{figure*}

\noindent \twocolumngrid 
\begin{widetext}
\vspace{-0.5cm}

\begin{equation}
\Psi_{E,j,j_{z}}^{+}\left(r,\Omega\right)=\mathcal{N}^{+}\left(\begin{array}{c}
\left[\cos\delta_{j}\left(E,\lambda b\right)J_{j}\left(\abs Er\right)-\text{sgn}\left(E\right)\sin\delta_{j}\left(E,\lambda b\right)Y_{j}\left(\abs Er\right)\right]\Theta_{j,j_{z}}^{-}\left(\Omega\right)\\
\iota\left[\cos\delta_{j}\left(E,\lambda b\right)J_{j+1}\left(\abs Er\right)-\text{sgn}\left(E\right)\sin\delta_{j}\left(E,\lambda b\right)Y_{j+1}\left(\abs Er\right)\right]\Theta_{j,j_{z}}^{+}\left(\Omega\right)
\end{array}\right)\label{eq:Psi+}
\end{equation}

\noindent and

\vspace{-0.35cm}

\begin{equation}
\Psi_{E,j,j_{z}}^{-}\left(r,\Omega\right)=\mathcal{N}^{-}\left(\begin{array}{c}
\left[\cos\delta_{j}\left(E,\lambda b\right)J_{j+1}\left(\abs Er\right)-\text{sgn}\left(E\right)\sin\delta_{j}\left(E,\lambda b\right)Y_{j+1}\left(\abs Er\right)\right]\Theta_{j,j_{z}}^{+}\left(\Omega\right)\\
-\iota\left[\cos\delta_{j}\left(E,\lambda b\right)J_{j}\left(\abs Er\right)-\text{sgn}\left(E\right)\sin\delta_{j}\left(E,\lambda b\right)Y_{j}\left(\abs Er\right)\right]\Theta_{j,j_{z}}^{-}\left(\Omega\right)
\end{array}\right),\label{eq:Psi-}
\end{equation}

\vspace{-0.35cm}
\end{widetext}

\noindent where $\mathcal{N}^{\pm}$ are complex normalization constants
and $\delta_{j}\left(E,\lambda b\right)$ are the energy-dependent
scattering phase shifts. From the forms of Eqs.\,(\ref{eq:Psi+})
and (\ref{eq:Psi-}), it is clear that adding $n\!\times\!\pi$ (with
integer $n$) to the phase shifts yields exactly the same spinor states,
apart from irrelevant global sign changes. Meanwhile, the scattering
phase shifts must always obey Eq.\,\eqref{eq:B/A-1-1-1-1}, which
guarantees continuity of the wavefunctions at $r\!=\!b$. However,
as explained in the main text, this condition does not uniquely define
the functions $\delta_{j}\!\left(\varepsilon,u\right)$, and a choice
must be made concerning the reference situation relative to which
the wave functions get dephased.

A natural choice is to define $\delta_{j}$ as the phase shift of
the wave function relative to the case when $u\!=\!0$. More precisely,
we can think of a situation in which the central potential is adiabatically
turned on and the instantaneous scattering eigenstates get progressively
more dephased at all energies. This convention is known to be a useful
one for Dirac fermions\,\citep{Ma85,Ma_2006}, and it can be achieved
by enforcing $\delta_{j}\left(\varepsilon\to\pm\infty,u\right)\to-u$.
It is important to remark that this choice is needed for us to relate
the change in the number of states inside a fixed spectral window
with the phase shifts of scattering states in that window. In Fig.\,\ref{fig:RadiusDependence_j1_2-1},
we depict the lowest-$j$ phase shifts, $\delta_{1/2}\!\left(\varepsilon,u\right)$
and $\delta_{3/2}\!\left(\varepsilon,u\right)$, as a function of
energy when $u$ is close to a critical value. These curves were obtained
using the previous convention for the phase shifts {[}$\delta_{j}\!\left(\varepsilon\!\to\!\pm\infty,u\right)\!\to\!-u${]},
which can be guaranteed by the following numerical integration:

$ $ 
\begin{widetext}
\vspace{-0.35cm}

\begin{equation}
\delta_{j}\left(\varepsilon,u\right)=\begin{cases}
-u+\int_{-\infty}^{\varepsilon}dx\frac{d}{dx}\arctan\left[\frac{\text{sgn}\left(x\right)J_{j}\left(\abs x\right)J_{j+1}\left(\abs{x-u}\right)-\text{sgn}\left(x-u\right)J_{j+1}\left(\abs x\right)J_{j}\left(\abs{x-u}\right)}{Y_{j}\left(\abs x\right)J_{j+1}\left(\abs{x-u}\right)-\text{sgn}\left(x\right)\text{sgn}\left(x-u\right)Y_{j+1}\left(\abs x\right)J_{j}\left(\abs{x-u}\right)}\right] & \text{if \ensuremath{\varepsilon<0}}\\
-u+\int_{\infty}^{\varepsilon}dx\frac{d}{dx}\arctan\left[\frac{\text{sgn}\left(x\right)J_{j}\left(\abs x\right)J_{j+1}\left(\abs{x-u}\right)-\text{sgn}\left(x-u\right)J_{j+1}\left(\abs x\right)J_{j}\left(\abs{x-u}\right)}{Y_{j}\left(\abs x\right)J_{j+1}\left(\abs{x-u}\right)-\text{sgn}\left(x\right)\text{sgn}\left(x-u\right)Y_{j+1}\left(\abs x\right)J_{j}\left(\abs{x-u}\right)}\right] & \text{if \ensuremath{\varepsilon\geq0}}
\end{cases}.
\end{equation}

\vspace{-0.35cm}
 
\end{widetext}

\noindent Defining $\delta_{j}$ by branches guarantees not only that
the appropriate asymptotic convention is obeyed but also that the
discontinuity due to zero-energy bound states is always avoided in
the integrals.

Finally, from Fig.\,\ref{fig:RadiusDependence_j1_2-1} it is clear
that a $\pi$ discontinuity develops at $\varepsilon\!=\!0$ when
the impurity parameter is critical. This is the trademark of a zero-energy
bound-state since, according to \textit{Levinson's theorem} for gapless
Dirac particles, the number of bound states with well-defined $j$,$j_{z}$,
and $\kappa$ is given as (see Ma and Ni\,\citep{Ma85})

\vspace{-0.5cm}

\begin{align}
n_{j,j_{z},\kappa=\pm}\left(u\right)\!\! & =\!\frac{1}{\pi}\left[\delta_{j}^{\pm}\left(0^{+},u\right)\!+\!\delta_{j}^{\pm}\left(0^{-},u\right)\right]\\
 & \mp\frac{\left(-1\right)^{j+1/2}}{2}\!\left[\sin^{2}\!\delta_{j}^{\pm}\!\left(0^{+}\!\!,u\right)\!-\!\sin^{2}\!\delta_{j}^{\pm}\!\left(0^{-}\!\!,u\right)\right],\nonumber 
\end{align}

\noindent which\,yields\,$n_{j,j_{z},\kappa=\pm}\!\left(u\!\neq\!n_{\text{c}}^{j}\right)\!=\!0$\,and\,$n_{j,j_{z},\kappa=\pm}\!\left(u\!=\!n_{\text{c}}^{j}\right)\!=\!1$.
This agrees with our earlier derivation of\,the\,zero-energy\,eigenstates\,in\,this\,system.

\vspace{-0.5cm}

\section{\label{sec:AppendixD}Additional Numerical Results and Technical
Details}

\subsubsection{Technical Description of the Numerical Method}

Here, we provide some technical details on the numerical method used
for calculating the density of states in the lattice model defined
in Eq.\,\eqref{eq:LatticeModel} of Sec.\,\ref{sec:Lattice-simulations}.
As explained there, the calculations used a kernel polynomial method
(KPM) \citep{Weise2006}, implemented in an efficient CPU parallelized
framework developed by some of us ($\texttt{QUANTUM\;Kite}\thinspace$\citep{Joao2020}).
We begin by outlining the basic elements of our numerical method.

Our aim is to calculate the intensive density of states (DoS) of a
finite quantum lattice system with $N$ degrees of freedom (in our
case, $N\!=\!4L^{3}$, as we have a simple cubic lattice with side
$L$ and four orbitals per site). This quantity is given generically
as $\rho\!\left(\varepsilon\right)d\varepsilon\!=\!\frac{1}{N}\!\sum_{\alpha}g_{\alpha}\delta\left(\varepsilon\!-\!\varepsilon_{\alpha}\right)d\varepsilon$,
where the summation is over eigenvalues of $\mathcal{H}$ and $g_{\alpha}$
is the degeneracy of each level. For our numerical purposes, $\rho\left(\varepsilon\right)d\varepsilon$
is expanded in Chebyshev polynomials of the kind, $T_{n}\!\left(x\right)$,
yielding

\vspace{-0.4cm}

\begin{align}
\rho_{\text{\!\!\ensuremath{\underset{\ensuremath{{\scriptscriptstyle K\!P\!M}}}{}}}}\!\!\!\!\left(\varepsilon,M\right)d\varepsilon\! & =\!\left\{ \!\frac{1}{\pi\sqrt{\lambda^{2}\!-\!\varepsilon^{2}}}\!\right.\!+\!2\!\sum_{n=1}^{M}\!\frac{g_{n}^{J}\!\left(M\right)T_{n}\!\left(\varepsilon/\lambda\right)}{\pi\sqrt{\lambda^{2}\!-\!\varepsilon^{2}}}\label{eq:DoSDefinition-1-1-1-1}\\
 & \qquad\qquad\qquad\qquad\qquad\left.\times\text{Tr}\left[T_{n}\left(\mathcal{\widetilde{H}}\right)\right]\right\} d\varepsilon,\nonumber 
\end{align}
where $M$ is a truncation order, $\mathcal{\widetilde{H}}\!=\!\mathcal{H}/\lambda$
is a rescaled Hamiltonian with spectrum contained inside the canonical
interval $\left[-1,1\right]$ and $\tilde{\varepsilon}\!=\!\varepsilon/\lambda$
is a rescaled energy. Also, $g_{n}^{J}\!\left(M\right)\!=\!\left[\!(M\!-\!n\!+\!1)\cos\!\left(\frac{\pi n}{M+1}\right)\!+\!\cot\!\left(\frac{\pi}{M+1}\right)\sin\!\left(\frac{\pi n}{M+1}\right)\right]\!/\!\left[M\!+\!1\right]$
is the so-called Jackson kernel which effectively damps the Gibbs
oscillations in the truncated approximation. This method introduces
a finite spectral resolution in the calculation which, near the band
center, is $\eta\!\left(M\right)\!\approx\!\pi\lambda/M$. The resolution
becomes narrower by increasing $M$.

Finally, we remark that given a function $f\!\left(x\right)$ which
is approximated with a finite resolution in $x$, the KPM-approximated
function is described as the following convolution integral:

\vspace{-0.6cm}

\begin{equation}
f_{\text{\!\!\ensuremath{\underset{\ensuremath{{\scriptscriptstyle K\!P\!M}}}{}}}}\!\!\!\!\!\left(x,\eta\right)\!=\!\!\int_{-1}^{1}\!\!\!\!\!d\tau f\left(x\right)\frac{e^{-\frac{\left(x-\tau\right)^{2}}{2\eta^{2}}}}{\sqrt{2\pi}\eta}.
\end{equation}
This result is used for most of the analysis done on our real-space
numerical results.

\vspace{-0.5cm}

\subsubsection{Lattice Model and Boundary Conditions}

\vspace{-0.2cm}

\noindent In this brief section, we provide details and illustrate
the lattice model used in all our numerical simulations. As referred
to in the main text, our basic lattice Hamiltonian $H^{\text{D}}$
was obtained by discretizing the continuum Dirac Hamiltonian {[}with
a scalar potential $\mathcal{U}\!\left(\mathbf{R}\right)${]}, $\mathcal{H}$,
in a simple cubic lattice with four orbitals per site. This tight-binding
model Hamiltonian reads

$ $

\begin{align}
H^{\text{D}} & =\frac{\iota v}{2a}\sum_{\mathbf{R}\in\mathcal{L}_{C}}\sum_{j=1}^{3}\left\{ \Psi_{\mathbf{R}}^{\dagger}\!\cdot\!\alpha^{j}\!\cdot\!\Psi_{\mathbf{R}+a\hat{e}_{j}}\!-\!\text{H.c.}\right\} \label{eq:Ham}\\
 & \qquad\qquad\qquad\qquad\qquad+\sum_{\mathbf{R}\in\mathcal{L}_{C}}\!\!\!\mathcal{U}\!\left(\mathbf{R}\right)\!\Psi_{\mathbf{R}}^{\dagger}\!\cdot\!\Psi_{\mathbf{R}},\nonumber 
\end{align}

\begin{figure*}
\begin{centering}
\includegraphics[scale=0.255]{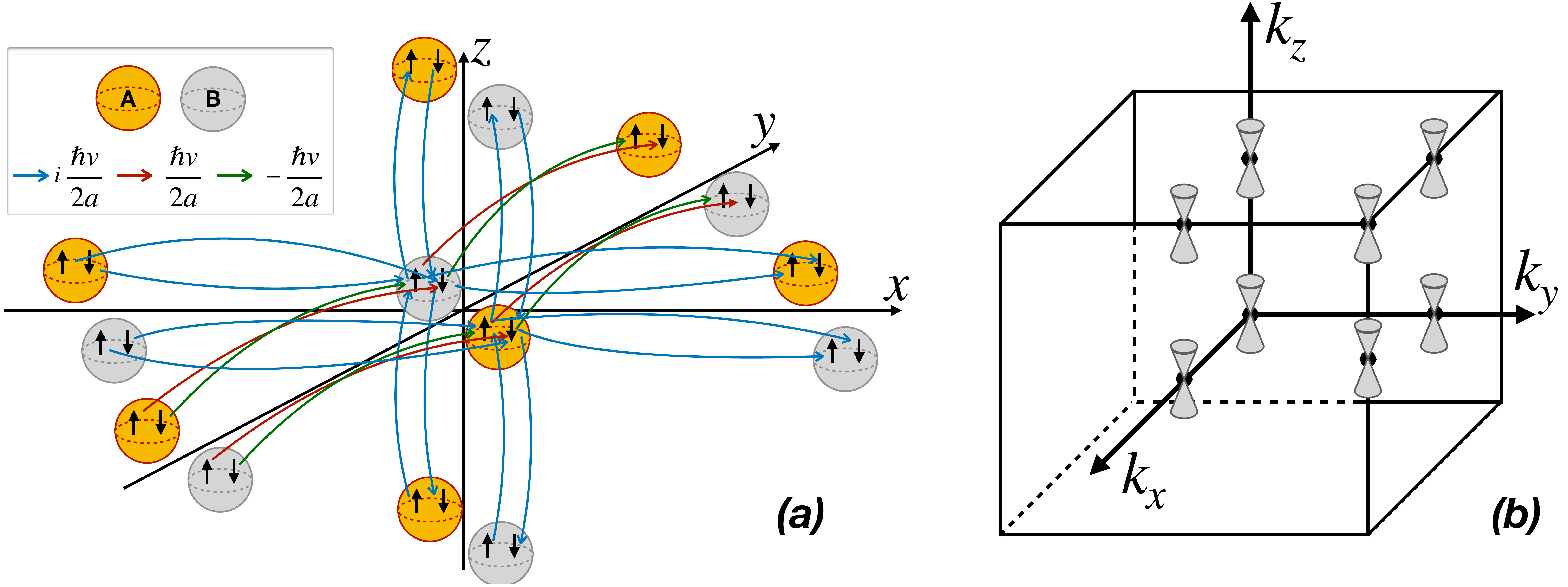} 
\par\end{centering}
\vspace{-0.3cm}

\caption{\label{fig:a)-Schematic-depiction}(a)\textbf{ }Schematic depiction
of the nearest-neighbor hoppings in the kinetic part of the Hamiltonian
$H^{\text{D}}$. Going in the direction inverse to that indicated
by the arrows means that the hopping will have the complex conjugate
value. (b) Representation of the simple cubic first Brillouin zone
of the model together with the places where the eight Dirac nodes
are present in the limit $\mathcal{U}\!\left(\mathbf{R}\right)\!=\!0$.\vspace{-0.9cm}
 }
\end{figure*}

\noindent \vspace{-0.3cm}

\noindent where $a$ is the lattice parameter and $\Psi_{\mathbf{R}}^{\dagger}\!=\!\left(c_{\mathbf{R},A,\uparrow}^{\dagger},c_{\mathbf{R},A,\downarrow}^{\dagger},c_{\mathbf{R},B,\uparrow}^{\dagger},c_{\mathbf{R},B,\downarrow}^{\dagger}\right)$
is a vector with onsite fermionic creation operators. Here, $A$ and
$B$ stand for two different sublattices while $\downarrow$ and $\uparrow$
are the two spin states in each orbital. Note that this convention
for naming the local single-particle states is consistent with the
previously defined intrinsic angular momentum operator $\mathbf{S}$.
In Fig.\,\ref{fig:a)-Schematic-depiction}(a), we depict this real-space
model in terms of its hoppings.

In the clean limit, $\mathcal{U}\!\left(\mathbf{R}\right)=0$, this
lattice model can be diagonalized by going to $\mathbf{k}$ space.
After doing that we obtain the particle-hole symmetric dispersion
relation

\vspace{-0.5cm}

\begin{equation}
E^{\text{c/v}}\!\left(\mathbf{k}\right)\!=\!\pm\frac{v}{a}\sqrt{\sin^{2}\!k_{x}a\!+\!\sin^{2}\!k_{y}a\!+\!\sin^{2}\!k_{z}a},
\end{equation}
where both the conduction and valence bands are twofold degenerate.
At half filling, this clearly reproduces a 3D Dirac semimetal, with
eight valleys placed at the time-reversal invariant momenta (TRIM)
of the first Brillouin zone. These are shown in Fig.\,\ref{fig:a)-Schematic-depiction}(b).
Near a TRIM, $\mathbf{K}_{\text{D}}$, the dispersion relation takes
the form

\vspace{-0.5cm}

\begin{equation}
E^{\text{c/v}}\!\left(\mathbf{k}\right)\!\approx\!\pm v\abs{\mathbf{k}\!-\!\mathbf{K}_{\text{D}}},
\end{equation}
which is exactly the same as we had in our original continuum Hamiltonian.
Nevertheless, the discretization of $\mathcal{H}_{0}$ introduces
a replication of the original four-fold degenerate Dirac cone into
eight disconnected ones.

Before ending this section, it is useful to calculate the normalized
DoS of the clean lattice model, as it is used explicitly in the analysis
of our numerical results. Using our previous definition, the intensive
DoS for this system (assuming a simulated lattice with $L^{3}$ sites)
reads

\vspace{-0.35cm}

\begin{equation}
\rho_{0}\!\left(\varepsilon\right)\!=\!\frac{2}{4L^{3}}\!\!\!\!\sum_{{\scriptscriptstyle \mathbf{k}\in\text{FBZ}}}\!\!\!\delta\!\left(\!\varepsilon\!\mp\!\frac{v}{a}\sqrt{\sin^{2}\!k_{x}a\!+\!\sin^{2}\!k_{y}a\!+\!\sin^{2}\!k_{z}a}\!\right).
\end{equation}

Due to particle-hole symmetry ($\varepsilon\!\to\!-\varepsilon$),
it suffices to evaluate de DoS at positive energies. Numerically,
we can choose a regular mesh in the first Brillouin zone (FBZ) of
the cubic lattice (equivalent to choosing a finite real-space cell)
and determine\,the\,normalized\,DoS.\,This\,is

\noindent \onecolumngrid

\begin{figure}[b]
\vspace{-0.3cm}

\begin{centering}
\hspace{-0.1cm}\includegraphics[scale=0.15]{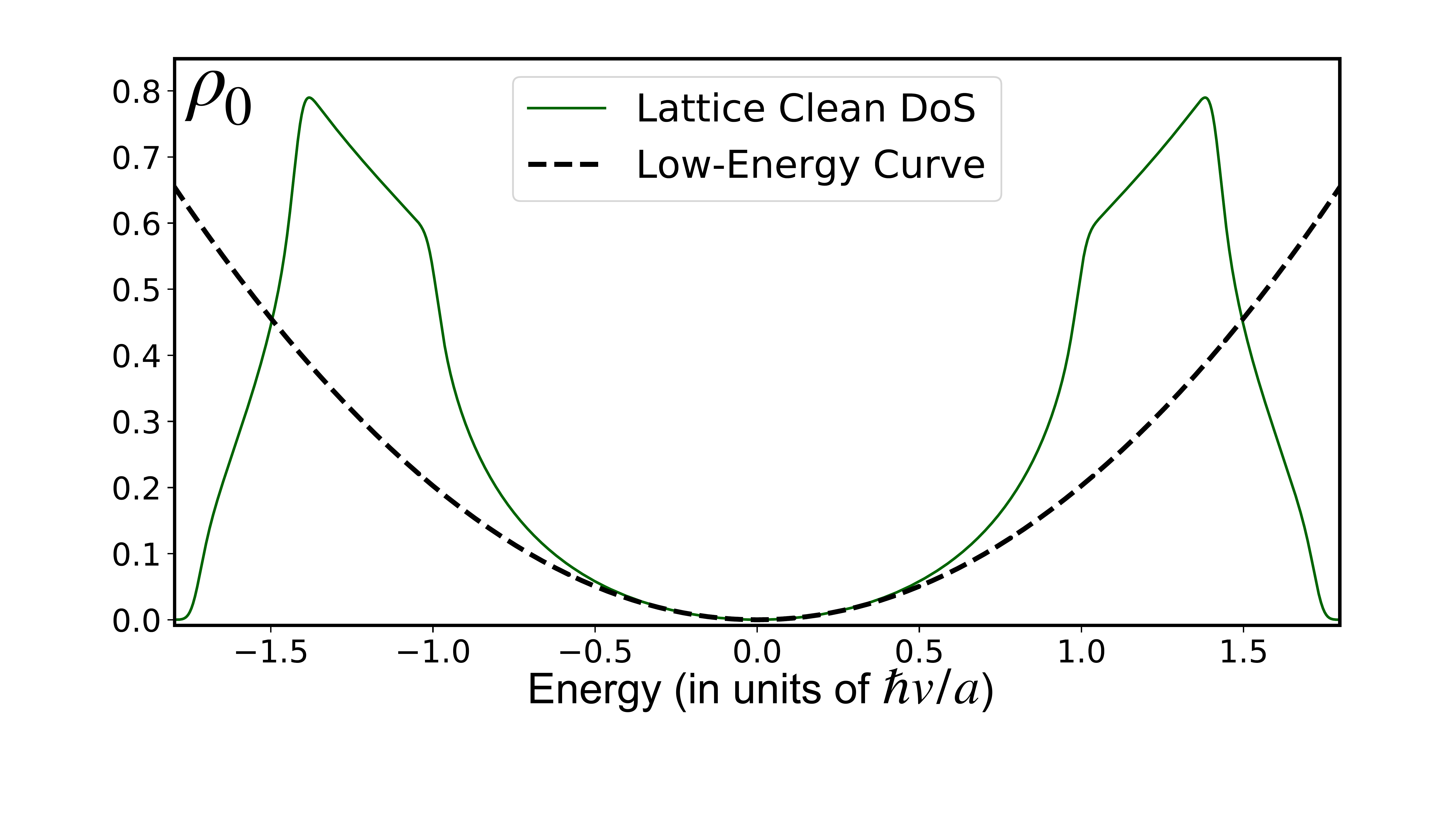} 
\par\end{centering}
\vspace{-0.3cm}

\caption{\textbf{\label{fig:a)-Schematic-depiction-1}}Plot of the normalized
density of states calculated for the lattice model of Eq.\,\eqref{eq:Ham}
(green curve). The dashed black curve corresponds to the quadratic
low-energy approximation to the DoS: $\rho_{0}\!\left(E\right)\!\simeq\!2E^{2}/\pi^{2}$.}
\end{figure}

\twocolumngrid

\noindent shown in Fig.\,\ref{fig:a)-Schematic-depiction-1}, together
with the corresponding low-energy quadratic approximation. The latter
is simply $\rho_{0}\left(E\right)\!=\!2E^{2}/\pi^{2}$, which is the
expression used in the main text.

\vspace{-0.35cm}

\subsubsection{Additional Results for the Resonances of a Single Sphere}

In this section, we present additional details on the numerical results
presented in the main\,text,\,together with additional results supporting
our conclusions. We begin by presenting the numerical results for
the change in the density of states due to a single extended sphere
in the center of a simulated supercell of size $L^{3}$. Despite simulating
a single sphere, we are actually sampling over random realizations
of boundary phase twists: This well-known technique helps the convergence
of the KPM calculations, by eliminating the mean-level spacing from
the problem. This method considers the computational domain as a supercell
that gets repeated in a periodic cubic superlattice. In the large-$L$
limit, periodicity artifacts eventually die out and fluctuations around
boundary-averaged values scale as $\propto\!L^{-3/2}$.

\begin{figure*}[t]
\vspace{-0.3cm}

\begin{centering}
\hspace{-0.5cm}\includegraphics[scale=0.23]{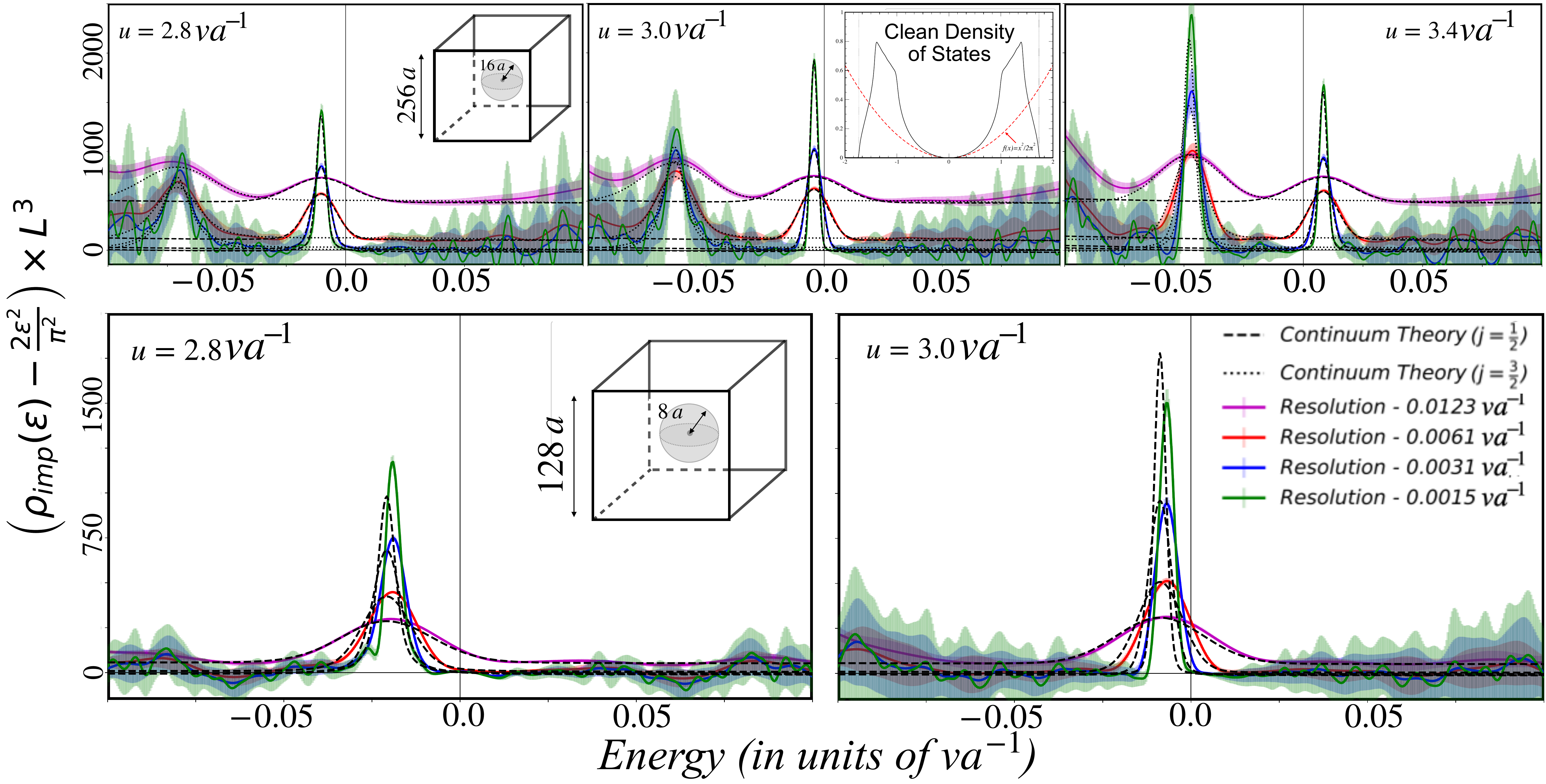} 
\par\end{centering}
\vspace{-0.4cm}

\caption{\label{fig:RadiusDependence_j1_2}\textbf{Top:} Plots of the change
in the density of states due to a single spherical impurity of strength
$u\!=\!2.8,3.0,3.4\;v/a$ and radius $b\!=\!16\,a$, inside a simulated
supercell of volume $256^{3}a^{3}$. The vertical widths of the numerical
curves are $95\%$ statistical error bars, with respect to the simultaneous
sampling over random KPM vectors and boundary phase twists. The agreement
with the resolution-corrected theoretical curves is perfect over the
entire range of resolutions used. \textbf{Bottom: }Results for a single
spherical impurity of strength $u\!=\!2.8,3.0\;v/a$ and radius $b\!=\!8\,a$,
inside a simulated supercell of volume $256^{3}\,a^{3}$. Finite-size
effects due to the discretization of the spherical impurity in the
lattice are now visible as a shift of the peak away from the node.}
\end{figure*}

Figure\,\ref{fig:RadiusDependence_j1_2} shows numerical results
for the $\Delta\rho_{\text{imp}}\!\left(E,u\right)\!L^{3}\!=\!\left(\rho_{\text{imp}}\!\left(E,u\right)\!-\!E^{2}/2\pi^{2}\right)L^{3}$
for values of $u$ around $\pi$, the first positive critical value
for a bound-state with $j\!=\!1/2$. These results are compared with
theoretical curves (dashed black lines), obtained from the result
of FSR,

\vspace{-0.6cm}

\begin{equation}
\Delta\rho_{\text{imp}}^{j}\!\left(E,u\right)\!=\!8\frac{2\left(2j\!+\!1\right)}{4\pi L^{3}a^{3}}\frac{d\delta_{j}\!\left(\varepsilon,u\right)}{d\varepsilon}\;\text{with}\:\varepsilon\!=\!Eb,\label{eq:FRS Result}
\end{equation}
convoluted with a Gaussian,

\vspace{-0.6cm}

\begin{align}
\Delta\tilde{\rho}_{\text{imp}}^{j}\!\left(E,u\right)\! & =\!-\frac{2E^{2}}{\pi^{2}}\!+\!\frac{1}{\sqrt{2\pi}\eta}\!\int_{-\infty}^{\infty}\!\!\!\!dxe^{-\frac{\left(E-x\right)^{2}}{2\eta^{2}}}\!\!\\
 & \qquad\qquad\qquad\qquad\times\left[\Delta\rho_{\text{imp}}^{j}\!\left(x,u\right)\!+\!\frac{2x^{2}}{\pi^{2}}\right]\nonumber 
\end{align}
to account for the finite spectral resolution ($\eta$) implied by
the numerical method. Note that Eq.\,\eqref{eq:FRS Result} includes
a factor of $8$ which accounts for the eight Dirac valleys existing
in our lattice model, as well as a $1/4$ due to the four orbitals
per site in our lattice model. The numerically calculated DoS is then
normalized by the number of states --- $4L^{3}$ --- and the clean
system has $\rho\!\left(E\right)\!\approx\!2E^{2}/\pi^{2}$ for $E\!\approx\!0$.

As can be seen from Fig.\,\ref{fig:RadiusDependence_j1_2}, the agreement
with the curves obtained from the continuum theory is perfect for
spheres of radius $b\!>\!16\,a$ with a concentration smaller than
$256^{-3}a^{-3}$ down to energy resolutions of $\text{meV}$. For
spheres of radius $b\!=\!8\;a$, one already observes deviations from
the continuum theory curves in the form of energy shifts (see bottom
panels in Fig.\,\ref{fig:RadiusDependence_j1_2}).

\begin{figure*}[t]
\vspace{-0.4cm}

\begin{centering}
\includegraphics[scale=0.23]{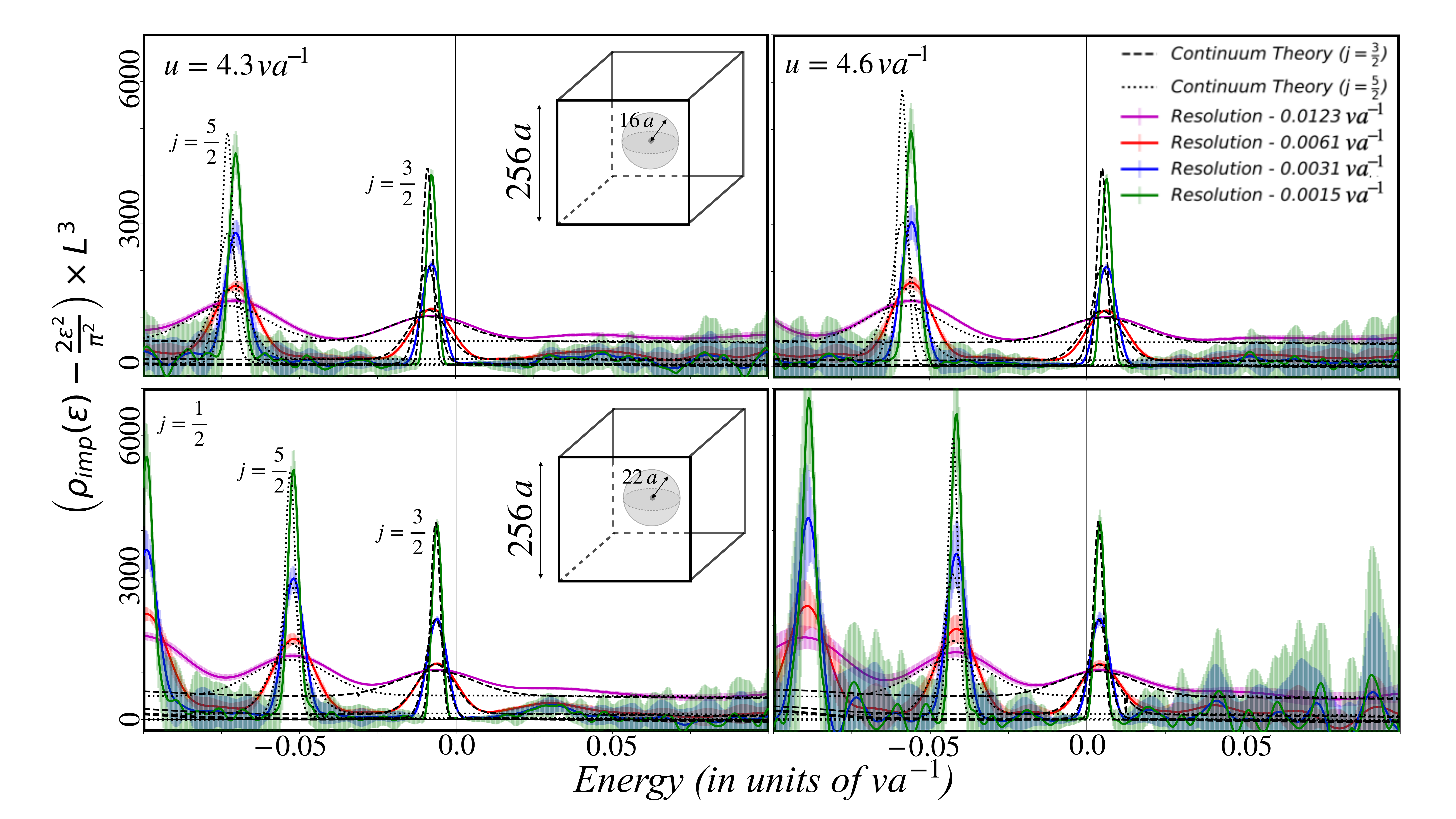} 
\par\end{centering}
\vspace{-0.4cm}

\caption{\label{fig:RadiusDependence_j1_2-2}\textbf{Top:} Plots of the change
in the density of states due to a single spherical impurity of strength
$u\!=\!4.3,4.6\;v/a$ and radius $b\!=\!16\,a$ inside a simulated
supercell of volume $256^{3}\,a^{3}$. The vertical widths of the
numerical curves are $95\%$ statistical error bars. The two visible
peaks correspond to the first resonances associated with $j\!=\!5/2$
and $j\!=\!3/2$, from left to right.\textbf{ Bottom}: Same calculations
done for an impurity of radius $b\!=\!22\,a$. The agreement with
the continuum theory is much better in this case.\vspace{-0.5cm}
 }
\end{figure*}

In Fig.\,\ref{fig:RadiusDependence_j1_2-2}, we represent analogous
high-energy-resolution numerical results for $u\!\approx\!4.493\cdots$
corresponding to the first resonance associated with $j\!\!=\!\!3/2$.
In the plots, one can also observe the next resonance (with $j\!=\!5/2$)
approaching the Dirac node. One can see that a radius of $16\,a$
is not sufficiently large to have a complete agreement between the
numerical peaks and the continuum theoretical curves.\,In the lower
panels, the calcula-

\noindent \onecolumngrid

\begin{figure*}
\begin{centering}
\includegraphics[scale=0.45]{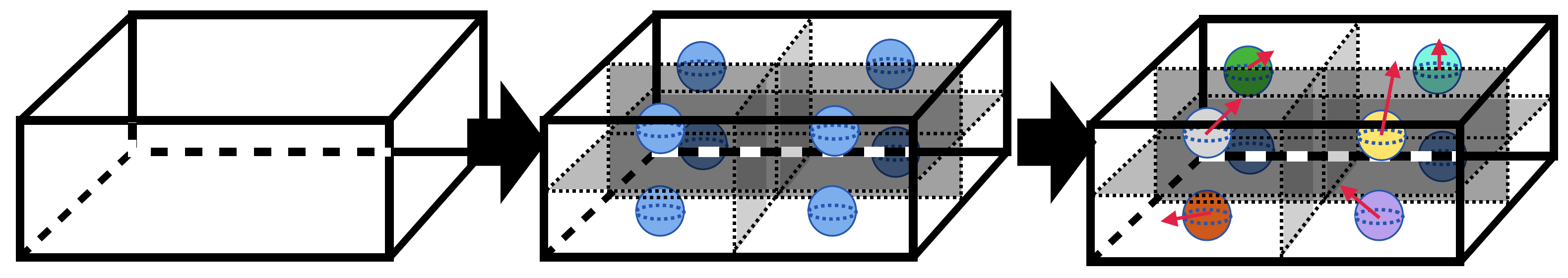} 
\par\end{centering}
\vspace{-0.3cm}

\caption{\label{fig:SchemeSeveralSpheres}Scheme of the procedure used to generate
a configuration of multiple random spheres inside the simulated supercell.\vspace{-0.5cm}
 }
\end{figure*}

\twocolumngrid

\clearpage{}

\noindent tion is repeated for a larger radius of the spherical impurity
($b\!=\!22\,a$), and a perfect agreement is then obtained for $j\!=\!3/2$.

Finally, it is important to analyze directly the case when the single
impurity inside the supercell is at a critical value. In this case,
we argue that uncoupled zero-energy eigenstates exist for the configuration,
contributing as $8\delta\!\left(E\right)\!/L^{3}$ to the DoS (contributions
coming from different valleys, as well as the factor of $4$ due to
the normalization to the total number of orbitals are included). One
can never see such a contribution numerically using the previous procedure,
but we can analyze its emergence as a function of the spectral resolution.
More precisely, we must have

\vspace{-0.65cm}

\begin{align}
\!\!\Delta\tilde{\rho}_{\text{imp}}^{j}\!\left(E,u_{\text{c}}\!\right)\! & =-\frac{2E^{2}}{\pi^{2}}\!+\!\frac{1}{\sqrt{2\pi}\eta}\!\int_{-\infty}^{\infty}\!\!\!\!dxe^{-\frac{\left(E-x\right)^{2}}{2\eta^{2}}}\\
 & \qquad\times\left[\!\frac{8}{L^{3}}\!\delta\!\left(x\right)\!+\!\frac{2x^{2}}{\pi^{2}}\!\right]\!=\!\frac{2\eta^{2}}{\pi^{2}}\!+\!\frac{8}{\sqrt{2\pi}L^{3}\eta}e^{-\frac{E^{2}}{2\eta^{2}}},\nonumber 
\end{align}
which is compared with numerical results (for $u\!=\!\pi$) in Fig.\,\ref{fig:LowEnergyDoS}.
The agreement is perfect.

To close this section, we remark that the main conclusions to be drawn
from the previous single-impurity results are threefold: (1) The continuum
theory describes the DoS peaks corresponding to resonances associated
with dilute \textit{near-critical} spherical impurities, provided
that this peak is located near the Dirac node (where the continuum
theory holds), the radius of the spheres is large enough, and the
distance between spheres is sufficiently large. (2) Larger-$j$ resonances
require the discretized spheres to be larger in order to reproduce
the continuum theory results for the same energy resolutions. (3)
Numerically, one can observe the emergence of a Dirac $\delta$ at
zero energy when the dilute impurities are all at critical values.

\vspace{-0.7cm}

\subsubsection{Details on the Simulation of the Average DoS for a System of Random
Impurities}

\vspace{-0.3cm}

Here, we provide details on the generation of the random distribution
of non-overlapping spheres in the lattice used to produce the numerical
results of Fig.\,\ref{fig:LowEnergyDoS}b. In order to do this, we
started by considering a simulated supercell (with twisted boundaries)
with $512^{3}$ unit cells ($\approx536\,000\,000$ orbitals), which
from the results of the previous single-impurity simulations is sufficient
to reproduce accurately the single sphere $\Delta\rho\!\left(E\right)$
at low energies if spheres of radius $16\,a$ are considered. Then,
we generate the potential associated with a regular cubic lattice
composed by the centers of such (discretized) spheres inside the simulated
cell. This procedure is equivalent to subdividing the original supercell
side by an integer number, generating a set of identical subcells.

Finally, each of the generated central points is randomly displaced
in three-dimensional space, and a potential strength is randomly chosen
for each impurity inside the supercell. This procedure guarantees
that there are no superpositions in any sample, as one restricts the
random displacement of the centers to keep it inside the corresponding
subcell.\,A schematic is depicted in Fig.\,\ref{fig:SchemeSeveralSpheres}.\,Once
the full potential landscape inside the supercell is created, the
remaining numerical procedure is identical to what was previously
described.

\bibliography{Refs}

\end{document}